\documentclass[prl,twocolumn,superscriptaddress,tightenlines,showpacs,longbibliography,floatfix,footinbib]{revtex4-1}

\pdfoutput=1

\usepackage{latexsym,amsmath,amsfonts,tabularx,amssymb,bm,empheq}
\usepackage{subfigure}

\usepackage{tensor}

\usepackage{blkarray}
\usepackage{mathtools}
\usepackage{multirow}

\begin{document}

\title{Random traction yielding transition in epithelial tissues}

\author{Aboutaleb Amiri}
\affiliation{Max Planck Institute for the Physics of Complex Systems, N\"othnitzer Str. 38,
01187 Dresden, Germany.}

\author{Charlie Duclut}
\affiliation{Max Planck Institute for the Physics of Complex Systems, N\"othnitzer Str. 38,
01187 Dresden, Germany.}
\affiliation{Universit\'e Paris Cit\'e, Laboratoire Mati\`ere et Syst\`emes Complexes (MSC), UMR 7057 CNRS, Paris,  France}
\affiliation{Laboratoire Physico-Chimie Curie, CNRS UMR 168, Institut Curie, Universit\'e PSL, Sorbonne Universit\'e, 75005, Paris, France}
\author{Frank J\"ulicher}
\affiliation{Max Planck Institute for the Physics of Complex Systems, N\"othnitzer Str. 38,
01187 Dresden, Germany.}
\affiliation{Cluster of Excellence Physics of Life, TU Dresden, Dresden, Germany}
\affiliation{Center for Systems Biology Dresden, Dresden, Germany}
\author{Marko Popovi\'c}
\affiliation{Max Planck Institute for the Physics of Complex Systems, N\"othnitzer Str. 38,
01187 Dresden, Germany.}
\affiliation{Center for Systems Biology Dresden, Dresden, Germany}

\date{\today}

\begin{abstract}
We investigate how randomly oriented cell traction forces lead to fluidisation in a vertex model of epithelial tissues. We find that the fluidisation occurs at a critical value of the traction force magnitude $F_c$. We show that this transition exhibits critical behaviour, similar to the yielding transition of sheared amorphous solids. However, we find that it belongs to a different universality class, even though it satisfies the same scaling relations between critical exponents established in the yielding transition of sheared amorphous solids. Our work provides a fluidisation mechanism through active force generation that could be relevant in biological tissues.
\end{abstract}

\maketitle

During tissue development, many cells collectively self-organize in dynamic patterns and morphologies. Therefore, a central problem in biophysics of development is understanding the interplay of tissue mechanics and active force generation~\cite{brodland2010video,keller2008forces,de2017forces,paul2013dpp,mayor2016front,cetera2018counter}. 
Cells in a tissue can generate traction forces through mechanical linkages with a substrate~\cite{sheetz1998cell,ingber1991integrins,du2005force,trepat2009physical} and impairment of this coupling can interrupt the movement of cells as observed, for example, in cancerous spheroid assays of carcinoma and human breast organoids~\cite{labernadie2017mechanically,cheung2013collective}. 
The response of biological tissues to mechanical forces is often described as that of viscoelastic active fluids~\cite{forgacs1998viscoelastic,iyer2019epithelial,lenne2022sculpting,duclut2021nonlinear}. However, recent experimental and theoretical studies have revealed complex mechanical phenomena, including jamming, glass transitions~\cite{berthier2013non,berthier2017active,bi2016motility,angelini2011glass,schoetz2013glassy,das2021controlled}, and yield stress rheology~\cite{mongera2018fluid,popovic2021inferring}. These observations suggest that developing biological tissues can behave as active amorphous solids.

Recently there has been an increasing interest in the rheology of active amorphous solids~\cite{Liao2018,Mandal2020}. In particular, comparing  uniform shear to random forcing of particles revealed a very similar non-linear response~\cite{morse2021direct}. 
A hallmark of sheared amorphous solids is a transition from a solid to a plastically flowing state at the yield stress $\Sigma_c$. The plastic strain rate $\dot{\gamma}$ at stress $\Sigma$ above the yielding transition typically follows the Herschel-Bulkley law $\dot{\gamma} \sim (\Sigma - \Sigma_c)^\beta$, where $\beta\geq 1$ is the flow exponent. Yielding has recently been reported under random forcing in systems of jammed self-propelled particles~\cite{villarroel2021critical}.
This raises the question of what is the nature of the yielding transition under random forces and how it is related to the yielding transition under uniform shear. Such random yielding is relevant in the context of biological tissues, allowing them to fluidise through generation of cell traction forces.

Here, we investigate the critical properties of the yielding under random traction forces using a vertex model of epithelial  tissues~\cite{farhadifar2007influence,popovic2021inferring}. Motivated by recent experiments on mouse pancreas spheres which suggest a presence of tissue fluidisation by cell traction forces~\cite{tan2022emergent}, we consider a vertex model with spherical geometry. This geometry is ubiquitous in multicellular systems such as the early developmental stages of many tissues, including early vertebrate embryos~\cite{kagawa2022human,valet2022mechanical}, and early stages of organoids~\cite{kim2020human,hsu2022activity,tan2022emergent}. 
We find that randomly oriented traction forces fluidise the cellular network beyond a critical magnitude $F_c$. 
We call this transition the \textit{random yielding transition} (RYT). We quantify the critical exponents characterizing overall cell flow, patterns of cell rearrangements, and even the geometry of the cellular network. 
We compare our results to the properties of the uniform shear yielding transition (YT). Interestingly, some critical exponents differ between the RYT and YT, implying that the transitions belong to different universality classes.
Furthermore, we find that RYT critical exponents satisfy the scaling relations between exponents established for the YT~\cite{lin2014density}. These relations imply that the statistical properties of tissue dynamics and cellular geometry are not independent.

\begin{figure}[b]
    \centering    \includegraphics[width=0.98\linewidth]{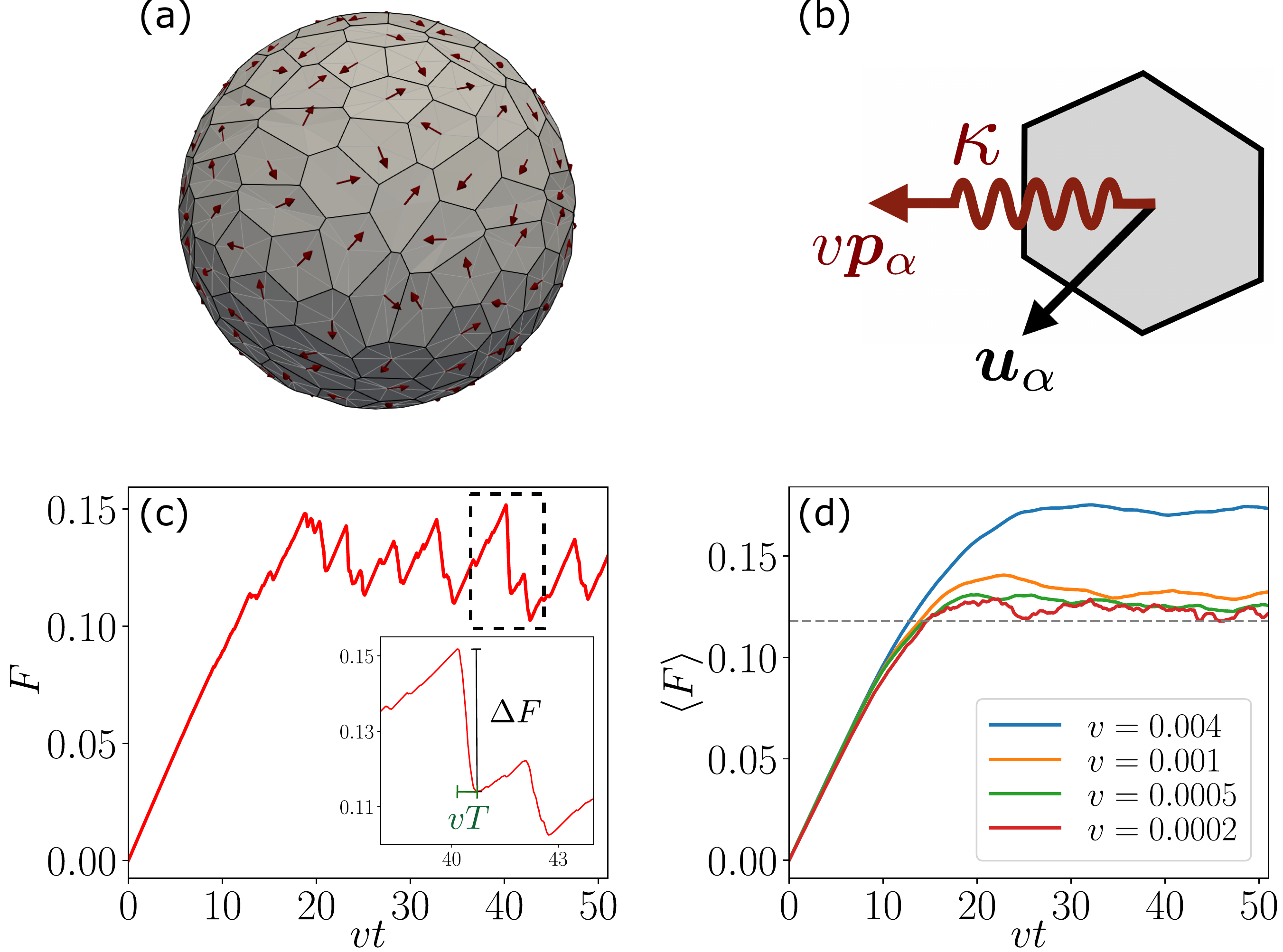}
    \caption{\textbf{(a)}~Spherical vertex model tissue with $N=200$ cells with randomly oriented traction forces (red arrows). \textbf{(b)}~Traction force is generated by extending a spring 
    of stiffness~$\kappa$, at speed~$v$ in direction of the polarity~${\bm p}_\alpha$. 
    \textbf{(c)}~Example of the tissue traction force magnitude dynamics $F$ as a function of spring displacement $v t$. \textbf{(d)}~Dynamics of ensemble-averaged tissue traction force magnitude. As the spring extension speed $v$ approaches the quasi-static driving limit $v\rightarrow0$, the traction force magnitude averaged over ensemble realisations $\langle F \rangle$ converges to its critical value $F_c$ marked by the dashed line (see SM).}
    \label{fig:Fig1}
    \end{figure}

\textbf{Random traction vertex model.} We extend the standard vertex model of epithelial tissues~\cite{farhadifar2007influence} to a spherical geometry (Fig.~\ref{fig:Fig1}a). 
We represent cells as polygons outlined by straight bonds, and constrain the polygon vertices to move on a sphere of radius $R$.
Geometry of the cellular network evolves following the dynamical equation:
\begin{align}
    \zeta \bm{u}_m = \bm{f}_m^{a} -\frac{\partial W}{\partial {\bm X}_m}
     + \bm{f}_m^{\textrm{n}}\, ,
    \label{eq_force_balance_vertex_model}
\end{align}
where $\bm{u}_m$ is the velocity of vertex $m$, $\zeta$ is the friction coefficient,  $\bm{f}_m^{a}$ is the traction force, $W$ is the vertex model energy function, and  $\bm{f}_m^{\textrm{n}}$ is the normal force constraining the motion of vertices on the sphere surface (see Supplemental Material~(SM) for details). 
The vertex model energy function that accounts for cell area elasticity and cell bond tension reads:
\begin{equation}
  W = \sum_{\alpha\in \text{cells}} \frac{1}{2} K\left( A^\alpha - A_0 \right)^2 
  +\sum_{\alpha \in \text{cells}} \frac{1}{2} \Lambda L^\alpha \, .
 \label{eq_tissue_work}
\end{equation}
Here, $A^\alpha$ is the cell area, $L^\alpha$ is the cell perimeter, $A_0$ is the preferred cell area, $K$ is the area stiffness, and $\Lambda$ is the perimeter tension magnitude~\cite{honda1984computer, farhadifar2007influence}.
We choose units of length, force and velocity to be $A_0^{1/2}$, $K A_0^{3/2}$ and $K A_0^{3/2}/\zeta$, respectively. The dimensionless bond tension $\overline{\Lambda}\equiv \Lambda/(K A_0^{3/2})$ is set to $\overline{\Lambda}= 0.1$. Implementation details are given in the SM.

We consider a planar cell polarity $\bm{p}_\alpha$ that directs the traction force exerted by cell $\alpha$ on the surrounding matrix (Fig.~\ref{fig:Fig1}b). We initialize the direction of the cell polarity vectors ${\bm p}_\alpha$ from a uniform distribution, and evolve it following  the dynamical equation:
\begin{equation}
\label{eq_dynamics_polarity_vertex_model}
    \frac{D{\bm{p}_\alpha}}{Dt} = 0 \quad ,
\end{equation}
where $D/Dt$ denotes a co-rotational time derivative (see SM), and we impose $|{\bm p}_\alpha|=1$ at each time.
We define the active traction force $\bm{f}_m^a$ on a vertex $m$ by uniformly redistributing the cell traction force $f_\alpha \bm{p}_\alpha$ of each of the abutting cells with $M_\alpha$ number of vertices: 
\begin{equation}
    \bm{f}_m^{a} = \sum_\alpha \frac{f_\alpha\bm{p}_\alpha}{M_\alpha}  \quad .
\end{equation}

\textbf{Random yielding transition.} Random traction forces induce stresses in the vertex model network. The stress magnitude is controlled by the magnitudes of cell traction forces $f_\alpha$. For small magnitudes of traction forces, we find that the elastic forces generated by the vertex model network balance the traction-induced forces, and the network remains solid. However, upon further increasing $f_\alpha$, the network begins to flow through cell rearrangements. To quantitatively explore this transition, we introduce the tissue traction force magnitude $F \equiv \sum_\alpha f_\alpha/N$, which in RYT plays the role analogous to the shear stress in the YT.

Application of uniform $f_\alpha$ is susceptible to finite-size effects that prevent us from probing the transition. Namely, a finite-size system can by chance reach an unusually stable configuration so that the system does not flow even at high $F$ values. To avoid this issue, we implement a model of traction forces where the attachment of a cell to the substrate moves with speed $v$ along the vector $\bm{p}_\alpha$ and the traction force is transmitted to a spring of stiffness $\kappa$ that connects the attachment and the cell (Fig.~\ref{fig:Fig1}b). Therefore, the dynamics of the traction force magnitude for a cell $\alpha$  follows:
\begin{equation}
\label{mixed_ensemble}
    \frac{d f_\alpha(t)}{dt} = - \kappa (\bm p_\alpha \cdot \bm u_\alpha - v) \quad .
\end{equation}
Here, the term $-\kappa \bm p_\alpha \cdot \bm u_\alpha$ represents the relaxation of the force in the spring due to motion of the cell with velocity~$\bm{u}_\alpha$. Limits of infinitely soft $\kappa \rightarrow 0$ and infinitely stiff $\kappa \rightarrow \infty$ springs correspond to imposed traction forces and imposed cell center velocities, respectively. 
In the following, we use~$\kappa=0.01$ and vary the imposed spring extension velocity~$v$.

An example of $F(t)$ dynamics as a function of spring displacement $v t$ is shown in Fig.~\ref{fig:Fig1}c (see also Movie~1). Initially, the cellular network responds elastically, and the traction forces grow linearly with spring displacement $v t$. As $F$ increases further, the cellular network begins to yield through cell rearrangements, visible as sharp drops of $F$ in Fig.~\ref{fig:Fig1}c. Finally, in the steady state, the system dynamics consist of periods of elastic loading punctuated by avalanches of cell rearrangements that are visible as sudden drops of $F$. Ensemble-averaged $F(t)$ for different values of $v$ is shown in  Fig.~\ref{fig:Fig1}d. 

The observed behaviour of $F$ is reminiscent of the stress \textit{vs} strain curve in sheared amorphous solids, such as metallic glasses~\cite{Sun2010}, where sudden drops of stress correspond to avalanches of particle rearrangements~\cite{Karmakar2010}. In amorphous solids near the YT, the avalanche size, defined as the number of particle rearrangements $S$ in an avalanche, is distributed according to a scaling law $P(S) = S^{-\tau} f(S/S_c)$, where $S_c$ is the cutoff beyond which $P(S)$ rapidly vanishes. The cutoff is set by the correlation length $\xi$: $S_c \sim \xi^{d_f}$, where $d_f$ is the avalanche fractal dimension~\cite{lin2014scaling}. However, approaching the YT, $\xi$ diverges and becomes larger than the system size. Therefore, in a finite system of $N$ cells, the cutoff $S_c$ is set by the system size $S_c \sim N^{d_f/d}$. Furthermore, the duration of an avalanche $T$ is expected to scale with the size as $T \sim S^{z/d_f}$, where $z$ is the dynamical exponent~\cite{lin2014scaling}. 

\begin{figure}
	\centering
    {\includegraphics[width=0.98\linewidth]{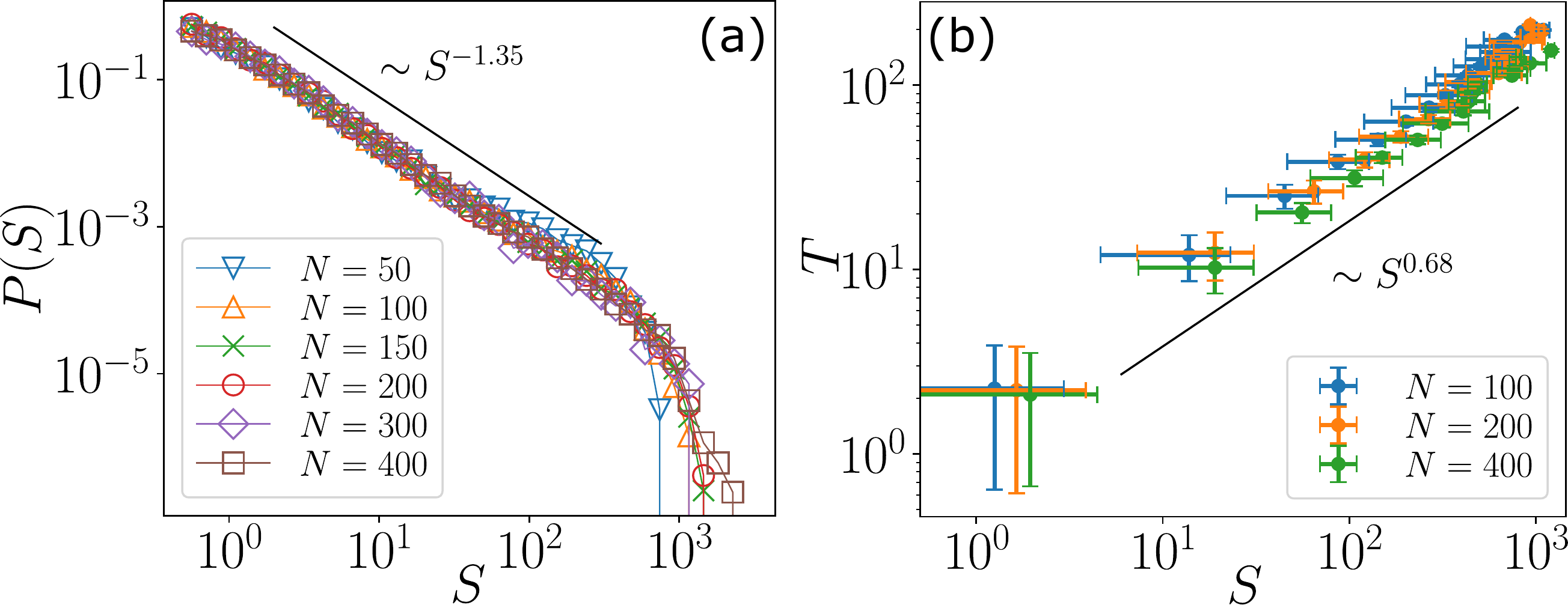}}
	\caption{%
    Avalanche statistics. \textbf{(a)}~Avalanche size has a power-law distribution $P(S)\sim S^{-\tau}$, with exponent $\tau=1.35\pm0.11$. \textbf{(b)}~Avalanche duration $T$ scales with avalanche size $S$ with exponent $z/d_f=0.68\pm0.04$. }
	\label{fig:T_S}
\end{figure}

To measure the avalanche size distribution, we measure drops in $F$ in the steady state at the lowest value $v= 2\cdot 10^{-4}$ we used. Then, we estimate the avalanche size corresponding to a force drop $\Delta F$ as $S\simeq N \Delta F/\kappa$. We find that the avalanche sizes are indeed power-law distributed, as shown in Fig.~\ref{fig:T_S}a, with a system-size dependent cutoff. Moreover, we find $\tau= 1.35 \pm 0.11$~\footnote{The uncertainty in our measurement of $\tau$ is mainly due to uncertainties in identification of avalanches at finite $v$, see SM.}, which is consistent with the values measured in the YT of $2d$ elastoplastic models ($\tau= 1.25 \pm 0.05$ in Ref.~\cite{talamali2011avalanches} and $\tau= 1.36 \pm 0.03$ in Ref.~\cite{lin2014scaling}), of a lattice model ($\tau= 1.342\pm 0.004$~\cite{Budrikis2013}) and of a finite element model ($\tau= 1.25\pm 0.05$~\cite{sandfeld2015avalanches}). 
We next estimated the avalanche fractal dimension $d_f= 0.75\pm 0.15$ by finite-size scaling analysis of the avalanche distribution cutoff using $S_c \sim \langle S^{3} \rangle / \langle S^2\rangle$, see SM. Finally, we find that the avalanche duration follows a power-law relationship with the avalanche size, see Fig.~\ref{fig:T_S}, from which we estimate $z/d_f=0.68 \pm 0.04$. 

To test whether the spherical geometry influences the exponent values, we also measured $\tau$, $d_f$ and $z/d_f$ in a bi-periodic $2d$ vertex model with identical random traction forces, and we found values of the exponents are consistent with the ones of the spherical model, see SM.

\begin{figure}
	\centering
    {\includegraphics[width=0.99\linewidth]{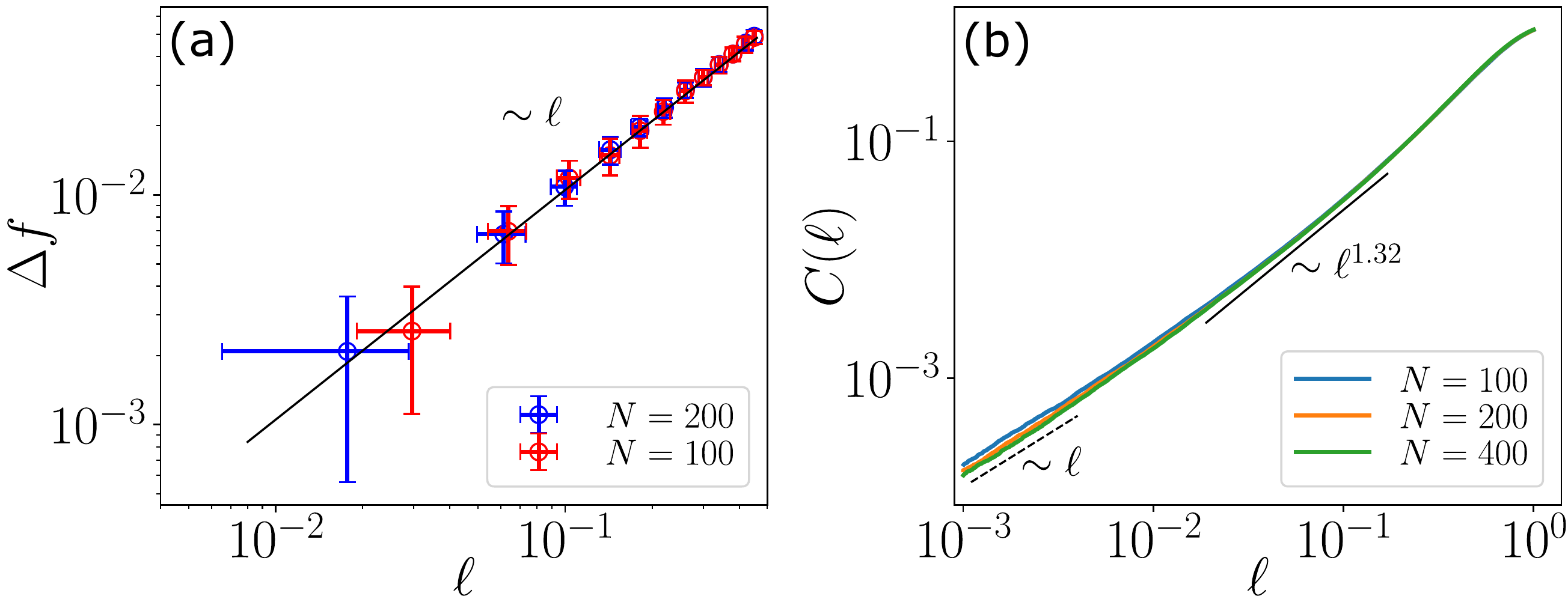}}
	\caption{%
    Density of plastic excitations in the tissue. \textbf{(a)}~Additional tension $\Delta f$ required to collapse the bond as a function of bond length $\ell$. A linear scaling is observed (solid line). \textbf{(b)}~Cumulative bond length distribution $C(\ell)$ in the steady state for $v=2\cdot 10^{-4}$. The predicted value of the exponent $\theta \approx 0.32$ is indicated by the solid line. At low $\ell$ we observe a linear scaling of $C(\ell)$ (dashed line), corresponding to a constant bond length distribution, as expected at finite $v$ and for finite system sizes.}
	\label{fig:P_l}
\end{figure}

\textbf{Scaling relations connect cellular dynamics and geometry.}  Exponents of YT are related through several scaling relations~\cite{lin2014scaling}. Here we examine two of these relations in the context of the RYT and show that in the vertex model with random traction forces they also provide a relationship between statistics of avalanches of cell rearrangements and cell bond length distribution.

The first scaling relation follows from the fact that in the steady state $\langle \Delta F\rangle = 0$~\cite{lin2014scaling}, which we now briefly reproduce. Increases of $F$ between avalanches are balanced by decreases during avalanches: $\langle |\Delta F| \rangle_{+} = \langle |\Delta F| \rangle_{-}$. The scaling of the average decrease of $F$ with system size can be estimated from the avalanche size distribution as $\langle |\Delta F|\rangle_- \sim \langle S \rangle / N \sim N^{(2 - \tau)d_f/d - 1}$. After an avalanche, $F$ will increase until the next T1 transition. Therefore, the increases in $F$ are determined by the network regions closest to a T1 transition. In amorphous solids the density of plastic excitations, defined as local increase in shear stress $\Delta \sigma$ required to trigger a plastic event, exhibits a pseudo-gap $P(\Delta \sigma) \sim \Delta \sigma^\theta$, with $\theta > 0$~\cite{Karmakar2010, lin2014density}. Thus, the average smallest $\Delta \sigma$ in a system of size $N$ scales as $\langle \Delta\sigma_\text{min}\rangle \sim N^{-1/(1 + \theta)}$ (see Ref.~\cite{lin2014scaling}). Since $\langle |\Delta F|\rangle_+ \sim \langle \Delta\sigma_\text{min}\rangle$ it follows that:
\begin{align}\label{eq:tau}
    \tau&= 2 - \frac{\theta}{1 + \theta}\frac{d}{d_f} \quad .
\end{align}
Using the measured values of $\tau$ and $d_f$, this scaling relation predicts $\theta=0.32\pm0.11$.

This prediction can be tested independently by considering the statistics of the bond length distribution as follows. In a vertex model network, each T1 transition corresponds to a vanishing bond;
hence, short bonds anticipate the upcoming T1 transitions. Due to cusps in the vertex model energy landscape at the onset of a T1, it was shown for the planar vertex model~\cite{popovic2021inferring} that the corresponding $\Delta \sigma$ is proportional to the bond length~$\ell$ of disappearing bonds. We show that this relation also holds in the spherical vertex model tissue, by measuring the additional tension $\Delta f$ required to shrink a bond of length $\ell$ to $0$, see Fig.~\ref{fig:P_l}a. In general, local change in shear stress $\Delta \sigma$ will generate a proportional change in the bond tension $\Delta f$. Therefore, observed scaling of imposed $\Delta f$ with bond length $\ell$ characterises the scaling of $\Delta \sigma$.
As a consequence, short bonds in the network for $F\leq F_c$ are distributed according to $P(\ell) \sim \ell^\theta$.  Figure~\ref{fig:P_l}b shows the cumulative bond length distribution $C(\ell)= \int_0^\ell P(\ell')d\ell'$ obtained in the steady-state simulation at $v= 2\cdot 10^{-4}$, where we measure bond lengths of networks at time points just after an avalanche. We find that the predicted value of the exponent $\theta$ is consistent with the bond length distribution (see also SM).
    
\begin{table*}[htb!]
\begin{center}
    \begin{tabular}{c|c|c|c}
         Exponent & Expression & RYT on a sphere & YT in $2d$ elastoplastic model \\
         \hline
         $\beta$ & $v\sim(\langle F\rangle-F_c)^{\beta}$ & $1.41\pm0.098$ & $1.52\pm0.05$\\
         $\tau$ & $P(S)\sim S^{-\tau}$ &  $1.35\pm0.11$ & $1.36\pm0.03$\\
         $z$ & $T\sim S^{z/{d_f}}$ & $0.51\pm0.11$ & $0.57\pm0.03$\\
         $d_f$ & $S_c\sim N^{{d_f}/d}$ & $0.75\pm0.15$ & $1.1\pm0.04$\\
         $\theta$ & $P(\Delta\sigma)\sim \Delta\sigma^\theta$ & $0.32\pm0.11$ & $0.57\pm0.01$
    \end{tabular}
 \caption{The critical exponents of RYT on a sphere in comparison with reported values for YT in a $2d$ elastoplastic model~\cite{lin2014scaling}.}    \label{table:exponents} 
\end{center}
\end{table*}

\begin{figure}
    \centering
    \includegraphics[width=0.65\linewidth]{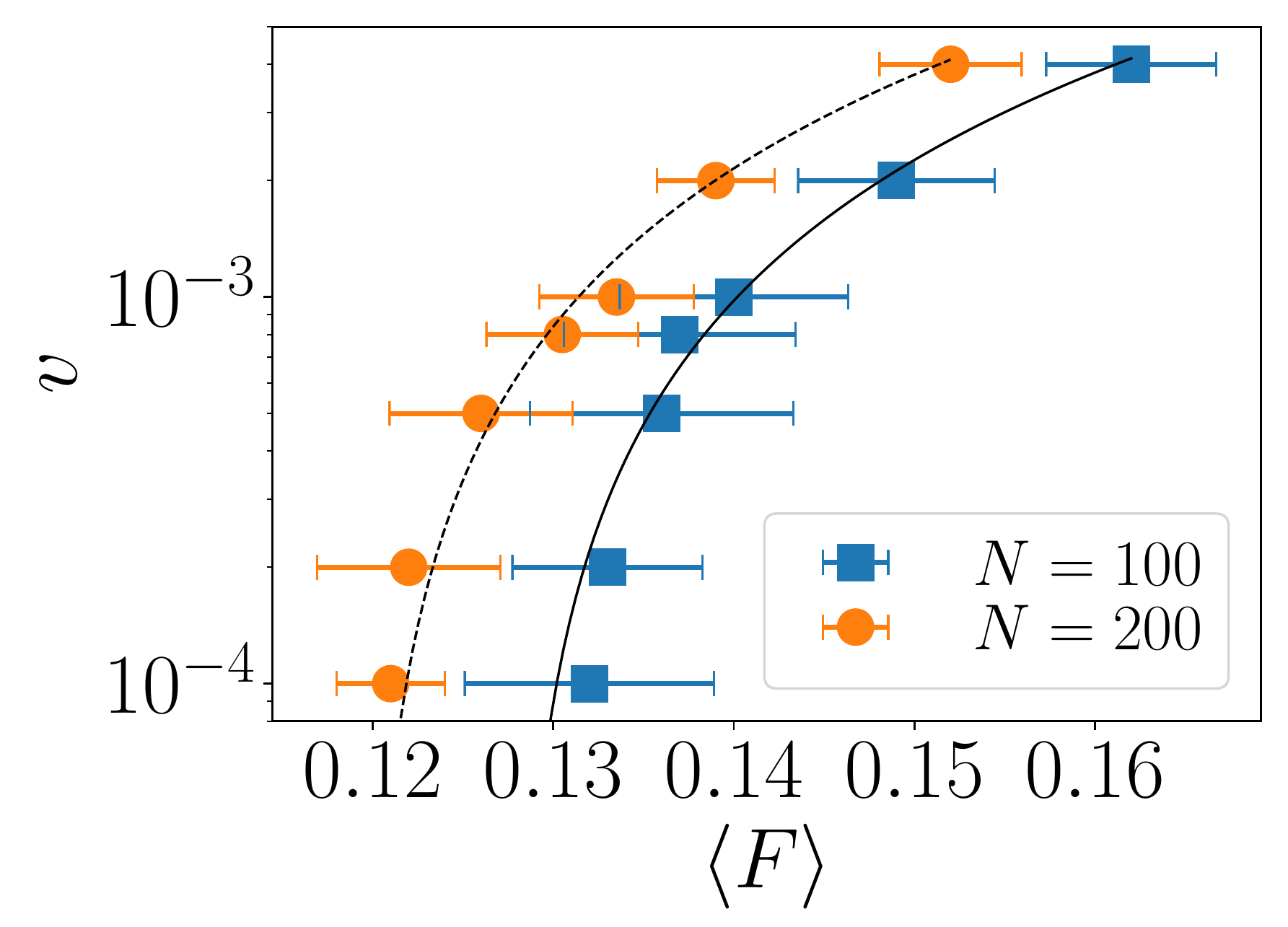}
    \caption{Steady-state flow curve measured in spherical vertex model networks with $N=100$ (blue squares) and $N=200$ (orange circles). Curves show best fit to $v\sim (\langle F\rangle -F_c)^{1.41}$ for $N=100$ (dashed line) and $N=200$ (solid line), see SM for discussion of $F_c$ finite size scaling.
    }
    \label{fig:HB_law}
    \end{figure}

The second scaling relation reflects that the flow in the vicinity of the critical point $F_c$ is composed of avalanches of spatial extension corresponding to the correlation length $\xi \sim (F - F_c)^{-\nu}$. Since the average avalanche size scales with $S \sim S_c \sim \xi^{d_f}$ and its duration scales as $T \sim \xi^z$ the contribution of the average avalanche to the overall flow $v$  will scale as $v \sim S/(T \xi^d) \sim (F - F_c)^{\nu (d - d_f + z)}$~\cite{lin2014scaling}. This determines the exponent $\beta= \nu(z + d - d_f)$ defined by $v \sim (F - F_c)^\beta$. Here we do not directly measure $\nu$ and instead we use an additional scaling relation $\nu = 1/(d - d_f)$
~\cite{lin2014scaling}. Therefore, we arrive at the relation:
\begin{align}\label{eq:scaling_relation_beta}
    \beta= 1 + \frac{z}{d - d_f} \quad,
\end{align}
which allows us to estimate  $\beta=1.41\pm0.098$.
To test this prediction, we analyze the steady-state flow properties for various magnitudes of loading rate $v$ as shown in Fig.~\ref{fig:HB_law} for two sizes $N= 100, 200$. We find a good agreement between numerical results and the value of $\beta$ predicted by the scaling relation~(\ref{eq:scaling_relation_beta}).

\textbf{Discussion.} 
We have described the critical properties of the RYT due to randomly oriented traction forces acting on a spherical epithelium. Our results show that this transition is closely related to the YT of sheared amorphous solids. Furthermore, we find that scaling relations constraining critical exponents of the YT also hold in the RYT, differing from the recent suggestion that one of the relations is violated~\cite{villarroel2021critical}. 
Furthermore, we independently measure the pseudo-gap exponent $\theta$ describing the density of plastic excitations. In our model, this exponent follows directly from the scaling of the distribution of cell bond lengths~\cite{popovic2021inferring} while it is typically difficult to access in particle models.

We find that the value of fractal dimension $d_f$ and pseudo-gap exponent $\theta$ are clearly different from the YT values; see Table~\ref{table:exponents}. In particular, $d_f \approx 1.1$ in $2d$ YT is associated with the one-dimensional shape of avalanches of plastic events, arising from the anisotropy of the Eshelby stress propagator of individual plastic events. In the RYT, the orientation of plastic events is not aligned, which breaks the preference of avalanches to occur along lines, and the value of $d_f = 0.75\pm 0.15$ smaller than $1$ shows that the spatial structure of avalanches is sparse. It is interesting to compare RYT to the yielding transition in a mean-field elastoplastic model where the Eshelby stress propagator is randomly redistributed in space, thereby removing all spatial correlations~\cite{lin2016mean}, where the pseudo-gap exponent $\theta = 0.39\pm 0.02$ has been reported numerically and supported by analytical calculations. This value is significantly lower than the $2d$ YT value $\theta \approx 0.57$~\cite{lin2014scaling}. However, since this is consistent with the value $\theta = 0.32\pm 0.11$ that we find in RYT, it would be interesting to test whether RYT is in the mean-field yielding transition universality class by carefully measuring the relevant critical exponents.

To test the influence of spherical geometry on the RYT we have measured the critical exponents $\tau$, $z$, and $d_f$ in flat $2d$ bi-periodic vertex model simulations (see SM). We found no significant difference in their values, which suggests that spherical geometry does not alter the critical behaviour of the vertex model near the RYT.

The dynamical exponent $z$ describes the dynamics of avalanche propagation $T \sim l^{z}$, where $l$ is the linear extension of the avalanche. The value $z= 0.51\pm 0.11$ we find is consistent with reported values in YT in $2d$ elastoplastic model $z= 0.57\pm 0.03$~\cite{lin2014scaling} and $z \simeq0.5$~\cite{jagla2017different}. 
However, in the thermodynamic limit $z < 1$ cannot hold due to the finite propagation speed of elastic interactions, which requires $z \geq 2$ in overdamped systems. Indeed $z \geq 2$ was reported in a large system of disks with overdamped dynamics~\cite{villarroel2021critical}. In finite systems $z$ can be smaller if the elastic interactions propagate through the system faster than the avalanches of cell rearrangements. Note that in elastoplastic models elastic interactions propagate instantaneously. The value of $z$ that we find suggests that, for the biologically relevant system sizes we consider, the elastic interactions in our model propagate much faster than avalanches, effectively behaving as instantaneous.   
 
The fluidisation through the generation of traction forces could allow the biological tissues to transition between a stable solid phase and a malleable fluid phase without the need to alter tissue density~\cite{mongera2018fluid} or cell mechanical properties~\cite{Bi2015}. We speculate a similar transition could occur in tissues where cells generate randomly oriented active stresses instead of traction forces.

\begin{acknowledgments}
We thank Matthieu Wyart for useful discussions.
This work was supported by the Federal Ministry of Education and Research (Bundesministerium f\"ur Bildung und Forschung, BMBF) under project 031L0160. CD acknowledges the support of a
postdoctoral fellowship from the LabEx ``Who Am I?'' (ANR-11-LABX-0071) and the Université Paris Cité IdEx (ANR-18-IDEX-0001) funded by the French Government through its ``Investments for the Future''. 
\end{acknowledgments}

\bibliography{biblio.bib}

\begin{thebibliography}{46}%
\makeatletter
\providecommand \@ifxundefined [1]{%
 \@ifx{#1\undefined}
}%
\providecommand \@ifnum [1]{%
 \ifnum #1\expandafter \@firstoftwo
 \else \expandafter \@secondoftwo
 \fi
}%
\providecommand \@ifx [1]{%
 \ifx #1\expandafter \@firstoftwo
 \else \expandafter \@secondoftwo
 \fi
}%
\providecommand \natexlab [1]{#1}%
\providecommand \enquote  [1]{``#1''}%
\providecommand \bibnamefont  [1]{#1}%
\providecommand \bibfnamefont [1]{#1}%
\providecommand \citenamefont [1]{#1}%
\providecommand \href@noop [0]{\@secondoftwo}%
\providecommand \href [0]{\begingroup \@sanitize@url \@href}%
\providecommand \@href[1]{\@@startlink{#1}\@@href}%
\providecommand \@@href[1]{\endgroup#1\@@endlink}%
\providecommand \@sanitize@url [0]{\catcode `\\12\catcode `\$12\catcode
  `\&12\catcode `\#12\catcode `\^12\catcode `\_12\catcode `\%12\relax}%
\providecommand \@@startlink[1]{}%
\providecommand \@@endlink[0]{}%
\providecommand \url  [0]{\begingroup\@sanitize@url \@url }%
\providecommand \@url [1]{\endgroup\@href {#1}{\urlprefix }}%
\providecommand \urlprefix  [0]{URL }%
\providecommand \Eprint [0]{\href }%
\providecommand \doibase [0]{http://dx.doi.org/}%
\providecommand \selectlanguage [0]{\@gobble}%
\providecommand \bibinfo  [0]{\@secondoftwo}%
\providecommand \bibfield  [0]{\@secondoftwo}%
\providecommand \translation [1]{[#1]}%
\providecommand \BibitemOpen [0]{}%
\providecommand \bibitemStop [0]{}%
\providecommand \bibitemNoStop [0]{.\EOS\space}%
\providecommand \EOS [0]{\spacefactor3000\relax}%
\providecommand \BibitemShut  [1]{\csname bibitem#1\endcsname}%
\let\auto@bib@innerbib\@empty
\bibitem [{\citenamefont {Brodland}\ \emph {et~al.}(2010)\citenamefont
  {Brodland}, \citenamefont {Conte}, \citenamefont {Cranston}, \citenamefont
  {Veldhuis}, \citenamefont {Narasimhan}, \citenamefont {Hutson}, \citenamefont
  {Jacinto}, \citenamefont {Ulrich}, \citenamefont {Baum},\ and\ \citenamefont
  {Miodownik}}]{brodland2010video}%
  \BibitemOpen
  \bibfield  {author} {\bibinfo {author} {\bibfnamefont {G~Wayne}\ \bibnamefont
  {Brodland}}, \bibinfo {author} {\bibfnamefont {Vito}\ \bibnamefont {Conte}},
  \bibinfo {author} {\bibfnamefont {P~Graham}\ \bibnamefont {Cranston}},
  \bibinfo {author} {\bibfnamefont {Jim}\ \bibnamefont {Veldhuis}}, \bibinfo
  {author} {\bibfnamefont {Sriram}\ \bibnamefont {Narasimhan}}, \bibinfo
  {author} {\bibfnamefont {M~Shane}\ \bibnamefont {Hutson}}, \bibinfo {author}
  {\bibfnamefont {Antonio}\ \bibnamefont {Jacinto}}, \bibinfo {author}
  {\bibfnamefont {Florian}\ \bibnamefont {Ulrich}}, \bibinfo {author}
  {\bibfnamefont {Buzz}\ \bibnamefont {Baum}}, \ and\ \bibinfo {author}
  {\bibfnamefont {Mark}\ \bibnamefont {Miodownik}},\ }\bibfield  {title}
  {\enquote {\bibinfo {title} {Video force microscopy reveals the mechanics of
  ventral furrow invagination in drosophila},}\ }\href@noop {} {\bibfield
  {journal} {\bibinfo  {journal} {Proceedings of the National Academy of
  Sciences}\ }\textbf {\bibinfo {volume} {107}},\ \bibinfo {pages}
  {22111--22116} (\bibinfo {year} {2010})}\BibitemShut {NoStop}%
\bibitem [{\citenamefont {Keller}\ \emph {et~al.}(2008)\citenamefont {Keller},
  \citenamefont {Shook},\ and\ \citenamefont {Skoglund}}]{keller2008forces}%
  \BibitemOpen
  \bibfield  {author} {\bibinfo {author} {\bibfnamefont {Ray}\ \bibnamefont
  {Keller}}, \bibinfo {author} {\bibfnamefont {David}\ \bibnamefont {Shook}}, \
  and\ \bibinfo {author} {\bibfnamefont {Paul}\ \bibnamefont {Skoglund}},\
  }\bibfield  {title} {\enquote {\bibinfo {title} {The forces that shape
  embryos: physical aspects of convergent extension by cell intercalation},}\
  }\href@noop {} {\bibfield  {journal} {\bibinfo  {journal} {Physical biology}\
  }\textbf {\bibinfo {volume} {5}},\ \bibinfo {pages} {015007} (\bibinfo {year}
  {2008})}\BibitemShut {NoStop}%
\bibitem [{\citenamefont {de~la Loza}\ and\ \citenamefont
  {Thompson}(2017)}]{de2017forces}%
  \BibitemOpen
  \bibfield  {author} {\bibinfo {author} {\bibfnamefont {MC~Diaz}\ \bibnamefont
  {de~la Loza}}\ and\ \bibinfo {author} {\bibfnamefont {BJ}~\bibnamefont
  {Thompson}},\ }\bibfield  {title} {\enquote {\bibinfo {title} {Forces shaping
  the drosophila wing},}\ }\href@noop {} {\bibfield  {journal} {\bibinfo
  {journal} {Mechanisms of Development}\ }\textbf {\bibinfo {volume} {144}},\
  \bibinfo {pages} {23--32} (\bibinfo {year} {2017})}\BibitemShut {NoStop}%
\bibitem [{\citenamefont {Paul}\ \emph {et~al.}(2013)\citenamefont {Paul},
  \citenamefont {Wang}, \citenamefont {Manivannan}, \citenamefont {Bonanno},
  \citenamefont {Lewis}, \citenamefont {Austin},\ and\ \citenamefont
  {Simcox}}]{paul2013dpp}%
  \BibitemOpen
  \bibfield  {author} {\bibinfo {author} {\bibfnamefont {Litty}\ \bibnamefont
  {Paul}}, \bibinfo {author} {\bibfnamefont {Shu-Huei}\ \bibnamefont {Wang}},
  \bibinfo {author} {\bibfnamefont {Sathiya~N}\ \bibnamefont {Manivannan}},
  \bibinfo {author} {\bibfnamefont {Liana}\ \bibnamefont {Bonanno}}, \bibinfo
  {author} {\bibfnamefont {Sarah}\ \bibnamefont {Lewis}}, \bibinfo {author}
  {\bibfnamefont {Christina~L}\ \bibnamefont {Austin}}, \ and\ \bibinfo
  {author} {\bibfnamefont {Amanda}\ \bibnamefont {Simcox}},\ }\bibfield
  {title} {\enquote {\bibinfo {title} {Dpp-induced egfr signaling triggers
  postembryonic wing development in drosophila},}\ }\href@noop {} {\bibfield
  {journal} {\bibinfo  {journal} {Proceedings of the National Academy of
  Sciences}\ }\textbf {\bibinfo {volume} {110}},\ \bibinfo {pages} {5058--5063}
  (\bibinfo {year} {2013})}\BibitemShut {NoStop}%
\bibitem [{\citenamefont {Mayor}\ and\ \citenamefont
  {Etienne-Manneville}(2016)}]{mayor2016front}%
  \BibitemOpen
  \bibfield  {author} {\bibinfo {author} {\bibfnamefont {Roberto}\ \bibnamefont
  {Mayor}}\ and\ \bibinfo {author} {\bibfnamefont {Sandrine}\ \bibnamefont
  {Etienne-Manneville}},\ }\bibfield  {title} {\enquote {\bibinfo {title} {The
  front and rear of collective cell migration},}\ }\href@noop {} {\bibfield
  {journal} {\bibinfo  {journal} {Nature reviews Molecular cell biology}\
  }\textbf {\bibinfo {volume} {17}},\ \bibinfo {pages} {97--109} (\bibinfo
  {year} {2016})}\BibitemShut {NoStop}%
\bibitem [{\citenamefont {Cetera}\ \emph {et~al.}(2018)\citenamefont {Cetera},
  \citenamefont {Leybova}, \citenamefont {Joyce},\ and\ \citenamefont
  {Devenport}}]{cetera2018counter}%
  \BibitemOpen
  \bibfield  {author} {\bibinfo {author} {\bibfnamefont {Maureen}\ \bibnamefont
  {Cetera}}, \bibinfo {author} {\bibfnamefont {Liliya}\ \bibnamefont
  {Leybova}}, \bibinfo {author} {\bibfnamefont {Bradley}\ \bibnamefont
  {Joyce}}, \ and\ \bibinfo {author} {\bibfnamefont {Danelle}\ \bibnamefont
  {Devenport}},\ }\bibfield  {title} {\enquote {\bibinfo {title}
  {Counter-rotational cell flows drive morphological and cell fate asymmetries
  in mammalian hair follicles},}\ }\href@noop {} {\bibfield  {journal}
  {\bibinfo  {journal} {Nature cell biology}\ }\textbf {\bibinfo {volume}
  {20}},\ \bibinfo {pages} {541--552} (\bibinfo {year} {2018})}\BibitemShut
  {NoStop}%
\bibitem [{\citenamefont {Sheetz}\ \emph {et~al.}(1998)\citenamefont {Sheetz},
  \citenamefont {Felsenfeld},\ and\ \citenamefont
  {Galbraith}}]{sheetz1998cell}%
  \BibitemOpen
  \bibfield  {author} {\bibinfo {author} {\bibfnamefont {Michael~P}\
  \bibnamefont {Sheetz}}, \bibinfo {author} {\bibfnamefont {Dan~P}\
  \bibnamefont {Felsenfeld}}, \ and\ \bibinfo {author} {\bibfnamefont
  {Catherine~G}\ \bibnamefont {Galbraith}},\ }\bibfield  {title} {\enquote
  {\bibinfo {title} {Cell migration: regulation of force on
  extracellular-matrix-integrin complexes},}\ }\href@noop {} {\bibfield
  {journal} {\bibinfo  {journal} {Trends in cell biology}\ }\textbf {\bibinfo
  {volume} {8}},\ \bibinfo {pages} {51--54} (\bibinfo {year}
  {1998})}\BibitemShut {NoStop}%
\bibitem [{\citenamefont {Ingber}(1991)}]{ingber1991integrins}%
  \BibitemOpen
  \bibfield  {author} {\bibinfo {author} {\bibfnamefont {Donald}\ \bibnamefont
  {Ingber}},\ }\bibfield  {title} {\enquote {\bibinfo {title} {Integrins as
  mechanochemical transducers},}\ }\href@noop {} {\bibfield  {journal}
  {\bibinfo  {journal} {Current opinion in cell biology}\ }\textbf {\bibinfo
  {volume} {3}},\ \bibinfo {pages} {841--848} (\bibinfo {year}
  {1991})}\BibitemShut {NoStop}%
\bibitem [{\citenamefont {Du~Roure}\ \emph {et~al.}(2005)\citenamefont
  {Du~Roure}, \citenamefont {Saez}, \citenamefont {Buguin}, \citenamefont
  {Austin}, \citenamefont {Chavrier}, \citenamefont {Siberzan},\ and\
  \citenamefont {Ladoux}}]{du2005force}%
  \BibitemOpen
  \bibfield  {author} {\bibinfo {author} {\bibfnamefont {Olivia}\ \bibnamefont
  {Du~Roure}}, \bibinfo {author} {\bibfnamefont {Alexandre}\ \bibnamefont
  {Saez}}, \bibinfo {author} {\bibfnamefont {Axel}\ \bibnamefont {Buguin}},
  \bibinfo {author} {\bibfnamefont {Robert~H}\ \bibnamefont {Austin}}, \bibinfo
  {author} {\bibfnamefont {Philippe}\ \bibnamefont {Chavrier}}, \bibinfo
  {author} {\bibfnamefont {Pascal}\ \bibnamefont {Siberzan}}, \ and\ \bibinfo
  {author} {\bibfnamefont {Benoit}\ \bibnamefont {Ladoux}},\ }\bibfield
  {title} {\enquote {\bibinfo {title} {Force mapping in epithelial cell
  migration},}\ }\href@noop {} {\bibfield  {journal} {\bibinfo  {journal}
  {Proceedings of the National Academy of Sciences}\ }\textbf {\bibinfo
  {volume} {102}},\ \bibinfo {pages} {2390--2395} (\bibinfo {year}
  {2005})}\BibitemShut {NoStop}%
\bibitem [{\citenamefont {Trepat}\ \emph {et~al.}(2009)\citenamefont {Trepat},
  \citenamefont {Wasserman}, \citenamefont {Angelini}, \citenamefont {Millet},
  \citenamefont {Weitz}, \citenamefont {Butler},\ and\ \citenamefont
  {Fredberg}}]{trepat2009physical}%
  \BibitemOpen
  \bibfield  {author} {\bibinfo {author} {\bibfnamefont {Xavier}\ \bibnamefont
  {Trepat}}, \bibinfo {author} {\bibfnamefont {Michael~R}\ \bibnamefont
  {Wasserman}}, \bibinfo {author} {\bibfnamefont {Thomas~E}\ \bibnamefont
  {Angelini}}, \bibinfo {author} {\bibfnamefont {Emil}\ \bibnamefont {Millet}},
  \bibinfo {author} {\bibfnamefont {David~A}\ \bibnamefont {Weitz}}, \bibinfo
  {author} {\bibfnamefont {James~P}\ \bibnamefont {Butler}}, \ and\ \bibinfo
  {author} {\bibfnamefont {Jeffrey~J}\ \bibnamefont {Fredberg}},\ }\bibfield
  {title} {\enquote {\bibinfo {title} {Physical forces during collective cell
  migration},}\ }\href@noop {} {\bibfield  {journal} {\bibinfo  {journal}
  {Nature physics}\ }\textbf {\bibinfo {volume} {5}},\ \bibinfo {pages}
  {426--430} (\bibinfo {year} {2009})}\BibitemShut {NoStop}%
\bibitem [{\citenamefont {Labernadie}\ \emph {et~al.}(2017)\citenamefont
  {Labernadie}, \citenamefont {Kato}, \citenamefont {Brugu{\'e}s},
  \citenamefont {Serra-Picamal}, \citenamefont {Derzsi}, \citenamefont
  {Arwert}, \citenamefont {Weston}, \citenamefont {Gonz{\'a}lez-Tarrag{\'o}},
  \citenamefont {Elosegui-Artola}, \citenamefont {Albertazzi} \emph
  {et~al.}}]{labernadie2017mechanically}%
  \BibitemOpen
  \bibfield  {author} {\bibinfo {author} {\bibfnamefont {Anna}\ \bibnamefont
  {Labernadie}}, \bibinfo {author} {\bibfnamefont {Takuya}\ \bibnamefont
  {Kato}}, \bibinfo {author} {\bibfnamefont {Agust{\'\i}}\ \bibnamefont
  {Brugu{\'e}s}}, \bibinfo {author} {\bibfnamefont {Xavier}\ \bibnamefont
  {Serra-Picamal}}, \bibinfo {author} {\bibfnamefont {Stefanie}\ \bibnamefont
  {Derzsi}}, \bibinfo {author} {\bibfnamefont {Esther}\ \bibnamefont {Arwert}},
  \bibinfo {author} {\bibfnamefont {Anne}\ \bibnamefont {Weston}}, \bibinfo
  {author} {\bibfnamefont {Victor}\ \bibnamefont {Gonz{\'a}lez-Tarrag{\'o}}},
  \bibinfo {author} {\bibfnamefont {Alberto}\ \bibnamefont {Elosegui-Artola}},
  \bibinfo {author} {\bibfnamefont {Lorenzo}\ \bibnamefont {Albertazzi}},
  \emph {et~al.},\ }\bibfield  {title} {\enquote {\bibinfo {title} {A
  mechanically active heterotypic e-cadherin/n-cadherin adhesion enables
  fibroblasts to drive cancer cell invasion},}\ }\href@noop {} {\bibfield
  {journal} {\bibinfo  {journal} {Nature cell biology}\ }\textbf {\bibinfo
  {volume} {19}},\ \bibinfo {pages} {224--237} (\bibinfo {year}
  {2017})}\BibitemShut {NoStop}%
\bibitem [{\citenamefont {Cheung}\ \emph {et~al.}(2013)\citenamefont {Cheung},
  \citenamefont {Gabrielson}, \citenamefont {Werb},\ and\ \citenamefont
  {Ewald}}]{cheung2013collective}%
  \BibitemOpen
  \bibfield  {author} {\bibinfo {author} {\bibfnamefont {Kevin~J}\ \bibnamefont
  {Cheung}}, \bibinfo {author} {\bibfnamefont {Edward}\ \bibnamefont
  {Gabrielson}}, \bibinfo {author} {\bibfnamefont {Zena}\ \bibnamefont {Werb}},
  \ and\ \bibinfo {author} {\bibfnamefont {Andrew~J}\ \bibnamefont {Ewald}},\
  }\bibfield  {title} {\enquote {\bibinfo {title} {Collective invasion in
  breast cancer requires a conserved basal epithelial program},}\ }\href@noop
  {} {\bibfield  {journal} {\bibinfo  {journal} {Cell}\ }\textbf {\bibinfo
  {volume} {155}},\ \bibinfo {pages} {1639--1651} (\bibinfo {year}
  {2013})}\BibitemShut {NoStop}%
\bibitem [{\citenamefont {Forgacs}\ \emph {et~al.}(1998)\citenamefont
  {Forgacs}, \citenamefont {Foty}, \citenamefont {Shafrir},\ and\ \citenamefont
  {Steinberg}}]{forgacs1998viscoelastic}%
  \BibitemOpen
  \bibfield  {author} {\bibinfo {author} {\bibfnamefont {Gabor}\ \bibnamefont
  {Forgacs}}, \bibinfo {author} {\bibfnamefont {Ramsey~A}\ \bibnamefont
  {Foty}}, \bibinfo {author} {\bibfnamefont {Yinon}\ \bibnamefont {Shafrir}}, \
  and\ \bibinfo {author} {\bibfnamefont {Malcolm~S}\ \bibnamefont
  {Steinberg}},\ }\bibfield  {title} {\enquote {\bibinfo {title} {Viscoelastic
  properties of living embryonic tissues: a quantitative study},}\ }\href@noop
  {} {\bibfield  {journal} {\bibinfo  {journal} {Biophysical journal}\ }\textbf
  {\bibinfo {volume} {74}},\ \bibinfo {pages} {2227--2234} (\bibinfo {year}
  {1998})}\BibitemShut {NoStop}%
\bibitem [{\citenamefont {Iyer}\ \emph {et~al.}(2019)\citenamefont {Iyer},
  \citenamefont {Piscitello-G{\'o}mez}, \citenamefont {Paijmans}, \citenamefont
  {J{\"u}licher},\ and\ \citenamefont {Eaton}}]{iyer2019epithelial}%
  \BibitemOpen
  \bibfield  {author} {\bibinfo {author} {\bibfnamefont {K~Venkatesan}\
  \bibnamefont {Iyer}}, \bibinfo {author} {\bibfnamefont {Romina}\ \bibnamefont
  {Piscitello-G{\'o}mez}}, \bibinfo {author} {\bibfnamefont {Joris}\
  \bibnamefont {Paijmans}}, \bibinfo {author} {\bibfnamefont {Frank}\
  \bibnamefont {J{\"u}licher}}, \ and\ \bibinfo {author} {\bibfnamefont
  {Suzanne}\ \bibnamefont {Eaton}},\ }\bibfield  {title} {\enquote {\bibinfo
  {title} {Epithelial viscoelasticity is regulated by mechanosensitive
  e-cadherin turnover},}\ }\href@noop {} {\bibfield  {journal} {\bibinfo
  {journal} {Current Biology}\ }\textbf {\bibinfo {volume} {29}},\ \bibinfo
  {pages} {578--591} (\bibinfo {year} {2019})}\BibitemShut {NoStop}%
\bibitem [{\citenamefont {Lenne}\ and\ \citenamefont
  {Trivedi}(2022)}]{lenne2022sculpting}%
  \BibitemOpen
  \bibfield  {author} {\bibinfo {author} {\bibfnamefont
  {Pierre-Fran{\c{c}}ois}\ \bibnamefont {Lenne}}\ and\ \bibinfo {author}
  {\bibfnamefont {Vikas}\ \bibnamefont {Trivedi}},\ }\bibfield  {title}
  {\enquote {\bibinfo {title} {Sculpting tissues by phase transitions},}\
  }\href@noop {} {\bibfield  {journal} {\bibinfo  {journal} {Nature
  Communications}\ }\textbf {\bibinfo {volume} {13}},\ \bibinfo {pages} {1--14}
  (\bibinfo {year} {2022})}\BibitemShut {NoStop}%
\bibitem [{\citenamefont {Duclut}\ \emph {et~al.}(2021)\citenamefont {Duclut},
  \citenamefont {Paijmans}, \citenamefont {Inamdar}, \citenamefont {Modes},\
  and\ \citenamefont {J{\"u}licher}}]{duclut2021nonlinear}%
  \BibitemOpen
  \bibfield  {author} {\bibinfo {author} {\bibfnamefont {Charlie}\ \bibnamefont
  {Duclut}}, \bibinfo {author} {\bibfnamefont {Joris}\ \bibnamefont
  {Paijmans}}, \bibinfo {author} {\bibfnamefont {Mandar~M}\ \bibnamefont
  {Inamdar}}, \bibinfo {author} {\bibfnamefont {Carl~D}\ \bibnamefont {Modes}},
  \ and\ \bibinfo {author} {\bibfnamefont {Frank}\ \bibnamefont
  {J{\"u}licher}},\ }\bibfield  {title} {\enquote {\bibinfo {title} {Nonlinear
  rheology of cellular networks},}\ }\href@noop {} {\bibfield  {journal}
  {\bibinfo  {journal} {Cells \& development}\ }\textbf {\bibinfo {volume}
  {168}},\ \bibinfo {pages} {203746} (\bibinfo {year} {2021})}\BibitemShut
  {NoStop}%
\bibitem [{\citenamefont {Berthier}\ and\ \citenamefont
  {Kurchan}(2013)}]{berthier2013non}%
  \BibitemOpen
  \bibfield  {author} {\bibinfo {author} {\bibfnamefont {Ludovic}\ \bibnamefont
  {Berthier}}\ and\ \bibinfo {author} {\bibfnamefont {Jorge}\ \bibnamefont
  {Kurchan}},\ }\bibfield  {title} {\enquote {\bibinfo {title} {Non-equilibrium
  glass transitions in driven and active matter},}\ }\href@noop {} {\bibfield
  {journal} {\bibinfo  {journal} {Nature Physics}\ }\textbf {\bibinfo {volume}
  {9}},\ \bibinfo {pages} {310--314} (\bibinfo {year} {2013})}\BibitemShut
  {NoStop}%
\bibitem [{\citenamefont {Berthier}\ \emph {et~al.}(2017)\citenamefont
  {Berthier}, \citenamefont {Flenner},\ and\ \citenamefont
  {Szamel}}]{berthier2017active}%
  \BibitemOpen
  \bibfield  {author} {\bibinfo {author} {\bibfnamefont {Ludovic}\ \bibnamefont
  {Berthier}}, \bibinfo {author} {\bibfnamefont {Elijah}\ \bibnamefont
  {Flenner}}, \ and\ \bibinfo {author} {\bibfnamefont {Grzegorz}\ \bibnamefont
  {Szamel}},\ }\bibfield  {title} {\enquote {\bibinfo {title} {How active
  forces influence nonequilibrium glass transitions},}\ }\href@noop {}
  {\bibfield  {journal} {\bibinfo  {journal} {New Journal of Physics}\ }\textbf
  {\bibinfo {volume} {19}},\ \bibinfo {pages} {125006} (\bibinfo {year}
  {2017})}\BibitemShut {NoStop}%
\bibitem [{\citenamefont {Bi}\ \emph {et~al.}(2016)\citenamefont {Bi},
  \citenamefont {Yang}, \citenamefont {Marchetti},\ and\ \citenamefont
  {Manning}}]{bi2016motility}%
  \BibitemOpen
  \bibfield  {author} {\bibinfo {author} {\bibfnamefont {Dapeng}\ \bibnamefont
  {Bi}}, \bibinfo {author} {\bibfnamefont {Xingbo}\ \bibnamefont {Yang}},
  \bibinfo {author} {\bibfnamefont {M~Cristina}\ \bibnamefont {Marchetti}}, \
  and\ \bibinfo {author} {\bibfnamefont {M~Lisa}\ \bibnamefont {Manning}},\
  }\bibfield  {title} {\enquote {\bibinfo {title} {Motility-driven glass and
  jamming transitions in biological tissues},}\ }\href@noop {} {\bibfield
  {journal} {\bibinfo  {journal} {Physical Review X}\ }\textbf {\bibinfo
  {volume} {6}},\ \bibinfo {pages} {021011} (\bibinfo {year}
  {2016})}\BibitemShut {NoStop}%
\bibitem [{\citenamefont {Angelini}\ \emph {et~al.}(2011)\citenamefont
  {Angelini}, \citenamefont {Hannezo}, \citenamefont {Trepat}, \citenamefont
  {Marquez}, \citenamefont {Fredberg},\ and\ \citenamefont
  {Weitz}}]{angelini2011glass}%
  \BibitemOpen
  \bibfield  {author} {\bibinfo {author} {\bibfnamefont {Thomas~E}\
  \bibnamefont {Angelini}}, \bibinfo {author} {\bibfnamefont {Edouard}\
  \bibnamefont {Hannezo}}, \bibinfo {author} {\bibfnamefont {Xavier}\
  \bibnamefont {Trepat}}, \bibinfo {author} {\bibfnamefont {Manuel}\
  \bibnamefont {Marquez}}, \bibinfo {author} {\bibfnamefont {Jeffrey~J}\
  \bibnamefont {Fredberg}}, \ and\ \bibinfo {author} {\bibfnamefont {David~A}\
  \bibnamefont {Weitz}},\ }\bibfield  {title} {\enquote {\bibinfo {title}
  {Glass-like dynamics of collective cell migration},}\ }\href@noop {}
  {\bibfield  {journal} {\bibinfo  {journal} {Proceedings of the National
  Academy of Sciences}\ }\textbf {\bibinfo {volume} {108}},\ \bibinfo {pages}
  {4714--4719} (\bibinfo {year} {2011})}\BibitemShut {NoStop}%
\bibitem [{\citenamefont {Schoetz}\ \emph {et~al.}(2013)\citenamefont
  {Schoetz}, \citenamefont {Lanio}, \citenamefont {Talbot},\ and\ \citenamefont
  {Manning}}]{schoetz2013glassy}%
  \BibitemOpen
  \bibfield  {author} {\bibinfo {author} {\bibfnamefont {Eva-Maria}\
  \bibnamefont {Schoetz}}, \bibinfo {author} {\bibfnamefont {Marcos}\
  \bibnamefont {Lanio}}, \bibinfo {author} {\bibfnamefont {Jared~A}\
  \bibnamefont {Talbot}}, \ and\ \bibinfo {author} {\bibfnamefont {M~Lisa}\
  \bibnamefont {Manning}},\ }\bibfield  {title} {\enquote {\bibinfo {title}
  {Glassy dynamics in three-dimensional embryonic tissues},}\ }\href@noop {}
  {\bibfield  {journal} {\bibinfo  {journal} {Journal of The Royal Society
  Interface}\ }\textbf {\bibinfo {volume} {10}},\ \bibinfo {pages} {20130726}
  (\bibinfo {year} {2013})}\BibitemShut {NoStop}%
\bibitem [{\citenamefont {Das}\ \emph {et~al.}(2021)\citenamefont {Das},
  \citenamefont {Sastry},\ and\ \citenamefont {Bi}}]{das2021controlled}%
  \BibitemOpen
  \bibfield  {author} {\bibinfo {author} {\bibfnamefont {Amit}\ \bibnamefont
  {Das}}, \bibinfo {author} {\bibfnamefont {Srikanth}\ \bibnamefont {Sastry}},
  \ and\ \bibinfo {author} {\bibfnamefont {Dapeng}\ \bibnamefont {Bi}},\
  }\bibfield  {title} {\enquote {\bibinfo {title} {Controlled neighbor
  exchanges drive glassy behavior, intermittency, and cell streaming in
  epithelial tissues},}\ }\href@noop {} {\bibfield  {journal} {\bibinfo
  {journal} {Physical Review X}\ }\textbf {\bibinfo {volume} {11}},\ \bibinfo
  {pages} {041037} (\bibinfo {year} {2021})}\BibitemShut {NoStop}%
\bibitem [{\citenamefont {Mongera}\ \emph {et~al.}(2018)\citenamefont
  {Mongera}, \citenamefont {Rowghanian}, \citenamefont {Gustafson},
  \citenamefont {Shelton}, \citenamefont {Kealhofer}, \citenamefont {Carn},
  \citenamefont {Serwane}, \citenamefont {Lucio}, \citenamefont {Giammona},\
  and\ \citenamefont {Camp{\`a}s}}]{mongera2018fluid}%
  \BibitemOpen
  \bibfield  {author} {\bibinfo {author} {\bibfnamefont {Alessandro}\
  \bibnamefont {Mongera}}, \bibinfo {author} {\bibfnamefont {Payam}\
  \bibnamefont {Rowghanian}}, \bibinfo {author} {\bibfnamefont {Hannah~J}\
  \bibnamefont {Gustafson}}, \bibinfo {author} {\bibfnamefont {Elijah}\
  \bibnamefont {Shelton}}, \bibinfo {author} {\bibfnamefont {David~A}\
  \bibnamefont {Kealhofer}}, \bibinfo {author} {\bibfnamefont {Emmet~K}\
  \bibnamefont {Carn}}, \bibinfo {author} {\bibfnamefont {Friedhelm}\
  \bibnamefont {Serwane}}, \bibinfo {author} {\bibfnamefont {Adam~A}\
  \bibnamefont {Lucio}}, \bibinfo {author} {\bibfnamefont {James}\ \bibnamefont
  {Giammona}}, \ and\ \bibinfo {author} {\bibfnamefont {Otger}\ \bibnamefont
  {Camp{\`a}s}},\ }\bibfield  {title} {\enquote {\bibinfo {title} {A
  fluid-to-solid jamming transition underlies vertebrate body axis
  elongation},}\ }\href@noop {} {\bibfield  {journal} {\bibinfo  {journal}
  {Nature}\ }\textbf {\bibinfo {volume} {561}},\ \bibinfo {pages} {401--405}
  (\bibinfo {year} {2018})}\BibitemShut {NoStop}%
\bibitem [{\citenamefont {Popovi{\'c}}\ \emph {et~al.}(2021)\citenamefont
  {Popovi{\'c}}, \citenamefont {Druelle}, \citenamefont {Dye}, \citenamefont
  {J{\"u}licher},\ and\ \citenamefont {Wyart}}]{popovic2021inferring}%
  \BibitemOpen
  \bibfield  {author} {\bibinfo {author} {\bibfnamefont {Marko}\ \bibnamefont
  {Popovi{\'c}}}, \bibinfo {author} {\bibfnamefont {Valentin}\ \bibnamefont
  {Druelle}}, \bibinfo {author} {\bibfnamefont {Natalie~A}\ \bibnamefont
  {Dye}}, \bibinfo {author} {\bibfnamefont {Frank}\ \bibnamefont
  {J{\"u}licher}}, \ and\ \bibinfo {author} {\bibfnamefont {Matthieu}\
  \bibnamefont {Wyart}},\ }\bibfield  {title} {\enquote {\bibinfo {title}
  {Inferring the flow properties of epithelial tissues from their geometry},}\
  }\href@noop {} {\bibfield  {journal} {\bibinfo  {journal} {New Journal of
  Physics}\ }\textbf {\bibinfo {volume} {23}},\ \bibinfo {pages} {033004}
  (\bibinfo {year} {2021})}\BibitemShut {NoStop}%
\bibitem [{\citenamefont {Liao}\ and\ \citenamefont {Xu}(2018)}]{Liao2018}%
  \BibitemOpen
  \bibfield  {author} {\bibinfo {author} {\bibfnamefont {Qinyi}\ \bibnamefont
  {Liao}}\ and\ \bibinfo {author} {\bibfnamefont {Ning}\ \bibnamefont {Xu}},\
  }\bibfield  {title} {\enquote {\bibinfo {title} {Criticality of the
  zero-temperature jamming transition probed by self-propelled particles},}\
  }\href@noop {} {\bibfield  {journal} {\bibinfo  {journal} {Soft Matter}\
  }\textbf {\bibinfo {volume} {14}},\ \bibinfo {pages} {853--860} (\bibinfo
  {year} {2018})}\BibitemShut {NoStop}%
\bibitem [{\citenamefont {Mandal}\ \emph {et~al.}(2020)\citenamefont {Mandal},
  \citenamefont {Bhuyan}, \citenamefont {Chaudhuri}, \citenamefont {Dasgupta},\
  and\ \citenamefont {Rao}}]{Mandal2020}%
  \BibitemOpen
  \bibfield  {author} {\bibinfo {author} {\bibfnamefont {Rituparno}\
  \bibnamefont {Mandal}}, \bibinfo {author} {\bibfnamefont {Pranab~Jyoti}\
  \bibnamefont {Bhuyan}}, \bibinfo {author} {\bibfnamefont {Pinaki}\
  \bibnamefont {Chaudhuri}}, \bibinfo {author} {\bibfnamefont {Chandan}\
  \bibnamefont {Dasgupta}}, \ and\ \bibinfo {author} {\bibfnamefont {Madan}\
  \bibnamefont {Rao}},\ }\bibfield  {title} {\enquote {\bibinfo {title}
  {Extreme active matter at high densities},}\ }\href@noop {} {\bibfield
  {journal} {\bibinfo  {journal} {Nature communications}\ }\textbf {\bibinfo
  {volume} {11}},\ \bibinfo {pages} {1--8} (\bibinfo {year}
  {2020})}\BibitemShut {NoStop}%
\bibitem [{\citenamefont {Morse}\ \emph {et~al.}(2021)\citenamefont {Morse},
  \citenamefont {Roy}, \citenamefont {Agoritsas}, \citenamefont {Stanifer},
  \citenamefont {Corwin},\ and\ \citenamefont {Manning}}]{morse2021direct}%
  \BibitemOpen
  \bibfield  {author} {\bibinfo {author} {\bibfnamefont {Peter~K}\ \bibnamefont
  {Morse}}, \bibinfo {author} {\bibfnamefont {Sudeshna}\ \bibnamefont {Roy}},
  \bibinfo {author} {\bibfnamefont {Elisabeth}\ \bibnamefont {Agoritsas}},
  \bibinfo {author} {\bibfnamefont {Ethan}\ \bibnamefont {Stanifer}}, \bibinfo
  {author} {\bibfnamefont {Eric~I}\ \bibnamefont {Corwin}}, \ and\ \bibinfo
  {author} {\bibfnamefont {M~Lisa}\ \bibnamefont {Manning}},\ }\bibfield
  {title} {\enquote {\bibinfo {title} {A direct link between active matter and
  sheared granular systems},}\ }\href@noop {} {\bibfield  {journal} {\bibinfo
  {journal} {Proceedings of the National Academy of Sciences}\ }\textbf
  {\bibinfo {volume} {118}} (\bibinfo {year} {2021})}\BibitemShut {NoStop}%
\bibitem [{\citenamefont {Villarroel}\ and\ \citenamefont
  {D{\"u}ring}(2021)}]{villarroel2021critical}%
  \BibitemOpen
  \bibfield  {author} {\bibinfo {author} {\bibfnamefont {Carlos}\ \bibnamefont
  {Villarroel}}\ and\ \bibinfo {author} {\bibfnamefont {Gustavo}\ \bibnamefont
  {D{\"u}ring}},\ }\bibfield  {title} {\enquote {\bibinfo {title} {Critical
  yielding rheology: from externally deformed glasses to active systems},}\
  }\href@noop {} {\bibfield  {journal} {\bibinfo  {journal} {Soft Matter}\
  }\textbf {\bibinfo {volume} {17}},\ \bibinfo {pages} {9944--9949} (\bibinfo
  {year} {2021})}\BibitemShut {NoStop}%
\bibitem [{\citenamefont {Farhadifar}\ \emph {et~al.}(2007)\citenamefont
  {Farhadifar}, \citenamefont {R{\"o}per}, \citenamefont {Aigouy},
  \citenamefont {Eaton},\ and\ \citenamefont
  {J{\"u}licher}}]{farhadifar2007influence}%
  \BibitemOpen
  \bibfield  {author} {\bibinfo {author} {\bibfnamefont {Reza}\ \bibnamefont
  {Farhadifar}}, \bibinfo {author} {\bibfnamefont {Jens-Christian}\
  \bibnamefont {R{\"o}per}}, \bibinfo {author} {\bibfnamefont {Benoit}\
  \bibnamefont {Aigouy}}, \bibinfo {author} {\bibfnamefont {Suzanne}\
  \bibnamefont {Eaton}}, \ and\ \bibinfo {author} {\bibfnamefont {Frank}\
  \bibnamefont {J{\"u}licher}},\ }\bibfield  {title} {\enquote {\bibinfo
  {title} {The influence of cell mechanics, cell-cell interactions, and
  proliferation on epithelial packing},}\ }\href@noop {} {\bibfield  {journal}
  {\bibinfo  {journal} {Current Biology}\ }\textbf {\bibinfo {volume} {17}},\
  \bibinfo {pages} {2095--2104} (\bibinfo {year} {2007})}\BibitemShut {NoStop}%
\bibitem [{\citenamefont {Tan}\ \emph {et~al.}(2022)\citenamefont {Tan},
  \citenamefont {Amiri}, \citenamefont {Seijo-Barandiar{\'a}n}, \citenamefont
  {Staddon}, \citenamefont {Materne}, \citenamefont {Tomas}, \citenamefont
  {Duclut}, \citenamefont {Popovi{\'c}}, \citenamefont {Grapin-Botton},\ and\
  \citenamefont {J{\"u}licher}}]{tan2022emergent}%
  \BibitemOpen
  \bibfield  {author} {\bibinfo {author} {\bibfnamefont {Tzer~Han}\
  \bibnamefont {Tan}}, \bibinfo {author} {\bibfnamefont {Aboutaleb}\
  \bibnamefont {Amiri}}, \bibinfo {author} {\bibfnamefont {Irene}\ \bibnamefont
  {Seijo-Barandiar{\'a}n}}, \bibinfo {author} {\bibfnamefont {Michael~F}\
  \bibnamefont {Staddon}}, \bibinfo {author} {\bibfnamefont {Anne}\
  \bibnamefont {Materne}}, \bibinfo {author} {\bibfnamefont {Sandra}\
  \bibnamefont {Tomas}}, \bibinfo {author} {\bibfnamefont {Charlie}\
  \bibnamefont {Duclut}}, \bibinfo {author} {\bibfnamefont {Marko}\
  \bibnamefont {Popovi{\'c}}}, \bibinfo {author} {\bibfnamefont {Anne}\
  \bibnamefont {Grapin-Botton}}, \ and\ \bibinfo {author} {\bibfnamefont
  {Frank}\ \bibnamefont {J{\"u}licher}},\ }\bibfield  {title} {\enquote
  {\bibinfo {title} {Emergent chirality in active solid rotation of pancreas
  spheres},}\ }\href@noop {} {\bibfield  {journal} {\bibinfo  {journal}
  {bioRxiv}\ } (\bibinfo {year} {2022})}\BibitemShut {NoStop}%
\bibitem [{\citenamefont {Kagawa}\ \emph {et~al.}(2022)\citenamefont {Kagawa},
  \citenamefont {Javali}, \citenamefont {Khoei}, \citenamefont {Sommer},
  \citenamefont {Sestini}, \citenamefont {Novatchkova}, \citenamefont
  {Scholte~op Reimer}, \citenamefont {Castel}, \citenamefont {Bruneau},
  \citenamefont {Maenhoudt} \emph {et~al.}}]{kagawa2022human}%
  \BibitemOpen
  \bibfield  {author} {\bibinfo {author} {\bibfnamefont {Harunobu}\
  \bibnamefont {Kagawa}}, \bibinfo {author} {\bibfnamefont {Alok}\ \bibnamefont
  {Javali}}, \bibinfo {author} {\bibfnamefont {Heidar~Heidari}\ \bibnamefont
  {Khoei}}, \bibinfo {author} {\bibfnamefont {Theresa~Maria}\ \bibnamefont
  {Sommer}}, \bibinfo {author} {\bibfnamefont {Giovanni}\ \bibnamefont
  {Sestini}}, \bibinfo {author} {\bibfnamefont {Maria}\ \bibnamefont
  {Novatchkova}}, \bibinfo {author} {\bibfnamefont {Yvonne}\ \bibnamefont
  {Scholte~op Reimer}}, \bibinfo {author} {\bibfnamefont {Ga{\"e}l}\
  \bibnamefont {Castel}}, \bibinfo {author} {\bibfnamefont {Alexandre}\
  \bibnamefont {Bruneau}}, \bibinfo {author} {\bibfnamefont {Nina}\
  \bibnamefont {Maenhoudt}},  \emph {et~al.},\ }\bibfield  {title} {\enquote
  {\bibinfo {title} {Human blastoids model blastocyst development and
  implantation},}\ }\href@noop {} {\bibfield  {journal} {\bibinfo  {journal}
  {Nature}\ }\textbf {\bibinfo {volume} {601}},\ \bibinfo {pages} {600--605}
  (\bibinfo {year} {2022})}\BibitemShut {NoStop}%
\bibitem [{\citenamefont {Valet}\ \emph {et~al.}(2022)\citenamefont {Valet},
  \citenamefont {Siggia},\ and\ \citenamefont
  {Brivanlou}}]{valet2022mechanical}%
  \BibitemOpen
  \bibfield  {author} {\bibinfo {author} {\bibfnamefont {Manon}\ \bibnamefont
  {Valet}}, \bibinfo {author} {\bibfnamefont {Eric~D}\ \bibnamefont {Siggia}},
  \ and\ \bibinfo {author} {\bibfnamefont {Ali~H}\ \bibnamefont {Brivanlou}},\
  }\bibfield  {title} {\enquote {\bibinfo {title} {Mechanical regulation of
  early vertebrate embryogenesis},}\ }\href@noop {} {\bibfield  {journal}
  {\bibinfo  {journal} {Nature Reviews Molecular Cell Biology}\ }\textbf
  {\bibinfo {volume} {23}},\ \bibinfo {pages} {169--184} (\bibinfo {year}
  {2022})}\BibitemShut {NoStop}%
\bibitem [{\citenamefont {Kim}\ \emph {et~al.}(2020)\citenamefont {Kim},
  \citenamefont {Koo},\ and\ \citenamefont {Knoblich}}]{kim2020human}%
  \BibitemOpen
  \bibfield  {author} {\bibinfo {author} {\bibfnamefont {Jihoon}\ \bibnamefont
  {Kim}}, \bibinfo {author} {\bibfnamefont {Bon-Kyoung}\ \bibnamefont {Koo}}, \
  and\ \bibinfo {author} {\bibfnamefont {Juergen~A}\ \bibnamefont {Knoblich}},\
  }\bibfield  {title} {\enquote {\bibinfo {title} {Human organoids: model
  systems for human biology and medicine},}\ }\href@noop {} {\bibfield
  {journal} {\bibinfo  {journal} {Nature Reviews Molecular Cell Biology}\
  }\textbf {\bibinfo {volume} {21}},\ \bibinfo {pages} {571--584} (\bibinfo
  {year} {2020})}\BibitemShut {NoStop}%
\bibitem [{\citenamefont {Hsu}\ \emph {et~al.}(2022)\citenamefont {Hsu},
  \citenamefont {Sciortino}, \citenamefont {de~la Trobe},\ and\ \citenamefont
  {Bausch}}]{hsu2022activity}%
  \BibitemOpen
  \bibfield  {author} {\bibinfo {author} {\bibfnamefont {Chiao-Peng}\
  \bibnamefont {Hsu}}, \bibinfo {author} {\bibfnamefont {Alfredo}\ \bibnamefont
  {Sciortino}}, \bibinfo {author} {\bibfnamefont {Yu~Alice}\ \bibnamefont
  {de~la Trobe}}, \ and\ \bibinfo {author} {\bibfnamefont {Andreas~R}\
  \bibnamefont {Bausch}},\ }\bibfield  {title} {\enquote {\bibinfo {title}
  {Activity-induced polar patterns of filaments gliding on a sphere},}\
  }\href@noop {} {\bibfield  {journal} {\bibinfo  {journal} {Nature
  communications}\ }\textbf {\bibinfo {volume} {13}},\ \bibinfo {pages} {1--8}
  (\bibinfo {year} {2022})}\BibitemShut {NoStop}%
\bibitem [{\citenamefont {Lin}\ \emph {et~al.}(2014{\natexlab{a}})\citenamefont
  {Lin}, \citenamefont {Saade}, \citenamefont {Lerner}, \citenamefont {Rosso},\
  and\ \citenamefont {Wyart}}]{lin2014density}%
  \BibitemOpen
  \bibfield  {author} {\bibinfo {author} {\bibfnamefont {Jie}\ \bibnamefont
  {Lin}}, \bibinfo {author} {\bibfnamefont {Alaa}\ \bibnamefont {Saade}},
  \bibinfo {author} {\bibfnamefont {Edan}\ \bibnamefont {Lerner}}, \bibinfo
  {author} {\bibfnamefont {Alberto}\ \bibnamefont {Rosso}}, \ and\ \bibinfo
  {author} {\bibfnamefont {Matthieu}\ \bibnamefont {Wyart}},\ }\bibfield
  {title} {\enquote {\bibinfo {title} {On the density of shear transformations
  in amorphous solids},}\ }\href@noop {} {\bibfield  {journal} {\bibinfo
  {journal} {EPL (Europhysics Letters)}\ }\textbf {\bibinfo {volume} {105}},\
  \bibinfo {pages} {26003} (\bibinfo {year} {2014}{\natexlab{a}})}\BibitemShut
  {NoStop}%
\bibitem [{\citenamefont {Honda}\ \emph {et~al.}(1984)\citenamefont {Honda},
  \citenamefont {Yamanaka},\ and\ \citenamefont
  {Dan-Sohkawa}}]{honda1984computer}%
  \BibitemOpen
  \bibfield  {author} {\bibinfo {author} {\bibfnamefont {H}~\bibnamefont
  {Honda}}, \bibinfo {author} {\bibfnamefont {H}~\bibnamefont {Yamanaka}}, \
  and\ \bibinfo {author} {\bibfnamefont {M}~\bibnamefont {Dan-Sohkawa}},\
  }\bibfield  {title} {\enquote {\bibinfo {title} {A computer simulation of
  geometrical configurations during cell division},}\ }\href@noop {} {\bibfield
   {journal} {\bibinfo  {journal} {Journal of theoretical biology}\ }\textbf
  {\bibinfo {volume} {106}},\ \bibinfo {pages} {423--435} (\bibinfo {year}
  {1984})}\BibitemShut {NoStop}%
\bibitem [{\citenamefont {Sun}\ \emph {et~al.}(2010)\citenamefont {Sun},
  \citenamefont {Yu}, \citenamefont {Jiao}, \citenamefont {Bai}, \citenamefont
  {Zhao},\ and\ \citenamefont {Wang}}]{Sun2010}%
  \BibitemOpen
  \bibfield  {author} {\bibinfo {author} {\bibfnamefont {B~A}\ \bibnamefont
  {Sun}}, \bibinfo {author} {\bibfnamefont {H~B}\ \bibnamefont {Yu}}, \bibinfo
  {author} {\bibfnamefont {W}~\bibnamefont {Jiao}}, \bibinfo {author}
  {\bibfnamefont {H~Y}\ \bibnamefont {Bai}}, \bibinfo {author} {\bibfnamefont
  {D~Q}\ \bibnamefont {Zhao}}, \ and\ \bibinfo {author} {\bibfnamefont {W~H}\
  \bibnamefont {Wang}},\ }\bibfield  {title} {\enquote {\bibinfo {title}
  {Plasticity of ductile metallic glasses: A self-organized critical state},}\
  }\href@noop {} {\bibfield  {journal} {\bibinfo  {journal} {Physical Review
  Letters}\ ,\ \bibinfo {pages} {4}} (\bibinfo {year} {2010})}\BibitemShut
  {NoStop}%
\bibitem [{\citenamefont {Karmakar}\ \emph {et~al.}(2010)\citenamefont
  {Karmakar}, \citenamefont {Lerner},\ and\ \citenamefont
  {Procaccia}}]{Karmakar2010}%
  \BibitemOpen
  \bibfield  {author} {\bibinfo {author} {\bibfnamefont {Smarajit}\
  \bibnamefont {Karmakar}}, \bibinfo {author} {\bibfnamefont {Edan}\
  \bibnamefont {Lerner}}, \ and\ \bibinfo {author} {\bibfnamefont {Itamar}\
  \bibnamefont {Procaccia}},\ }\bibfield  {title} {\enquote {\bibinfo {title}
  {Statistical physics of the yielding transition in amorphous solids},}\
  }\href@noop {} {\bibfield  {journal} {\bibinfo  {journal} {Physical Review
  E}\ }\textbf {\bibinfo {volume} {82}},\ \bibinfo {pages} {055103} (\bibinfo
  {year} {2010})}\BibitemShut {NoStop}%
\bibitem [{\citenamefont {Lin}\ \emph {et~al.}(2014{\natexlab{b}})\citenamefont
  {Lin}, \citenamefont {Lerner}, \citenamefont {Rosso},\ and\ \citenamefont
  {Wyart}}]{lin2014scaling}%
  \BibitemOpen
  \bibfield  {author} {\bibinfo {author} {\bibfnamefont {Jie}\ \bibnamefont
  {Lin}}, \bibinfo {author} {\bibfnamefont {Edan}\ \bibnamefont {Lerner}},
  \bibinfo {author} {\bibfnamefont {Alberto}\ \bibnamefont {Rosso}}, \ and\
  \bibinfo {author} {\bibfnamefont {Matthieu}\ \bibnamefont {Wyart}},\
  }\bibfield  {title} {\enquote {\bibinfo {title} {Scaling description of the
  yielding transition in soft amorphous solids at zero temperature},}\
  }\href@noop {} {\bibfield  {journal} {\bibinfo  {journal} {Proceedings of the
  National Academy of Sciences}\ }\textbf {\bibinfo {volume} {111}},\ \bibinfo
  {pages} {14382--14387} (\bibinfo {year} {2014}{\natexlab{b}})}\BibitemShut
  {NoStop}%
\bibitem [{Note1()}]{Note1}%
  \BibitemOpen
  \bibinfo {note} {The uncertainty in our measurement of $\tau $ is mainly due
  to uncertainties in identification of avalanches at finite $v$, see
  SM.}\BibitemShut {Stop}%
\bibitem [{\citenamefont {Talamali}\ \emph {et~al.}(2011)\citenamefont
  {Talamali}, \citenamefont {Pet{\"a}j{\"a}}, \citenamefont {Vandembroucq},\
  and\ \citenamefont {Roux}}]{talamali2011avalanches}%
  \BibitemOpen
  \bibfield  {author} {\bibinfo {author} {\bibfnamefont {Mehdi}\ \bibnamefont
  {Talamali}}, \bibinfo {author} {\bibfnamefont {Viljo}\ \bibnamefont
  {Pet{\"a}j{\"a}}}, \bibinfo {author} {\bibfnamefont {Damien}\ \bibnamefont
  {Vandembroucq}}, \ and\ \bibinfo {author} {\bibfnamefont {St{\'e}phane}\
  \bibnamefont {Roux}},\ }\bibfield  {title} {\enquote {\bibinfo {title}
  {Avalanches, precursors, and finite-size fluctuations in a mesoscopic model
  of amorphous plasticity},}\ }\href@noop {} {\bibfield  {journal} {\bibinfo
  {journal} {Physical Review E}\ }\textbf {\bibinfo {volume} {84}},\ \bibinfo
  {pages} {016115} (\bibinfo {year} {2011})}\BibitemShut {NoStop}%
\bibitem [{\citenamefont {Budrikis}\ and\ \citenamefont
  {Zapperi}(2013)}]{Budrikis2013}%
  \BibitemOpen
  \bibfield  {author} {\bibinfo {author} {\bibfnamefont {Zoe}\ \bibnamefont
  {Budrikis}}\ and\ \bibinfo {author} {\bibfnamefont {Stefano}\ \bibnamefont
  {Zapperi}},\ }\bibfield  {title} {\enquote {\bibinfo {title} {Avalanche
  localization and crossover scaling in amorphous plasticity},}\ }\href@noop {}
  {\bibfield  {journal} {\bibinfo  {journal} {Physical Review E}\ ,\ \bibinfo
  {pages} {7}} (\bibinfo {year} {2013})}\BibitemShut {NoStop}%
\bibitem [{\citenamefont {Sandfeld}\ \emph {et~al.}(2015)\citenamefont
  {Sandfeld}, \citenamefont {Budrikis}, \citenamefont {Zapperi},\ and\
  \citenamefont {Castellanos}}]{sandfeld2015avalanches}%
  \BibitemOpen
  \bibfield  {author} {\bibinfo {author} {\bibfnamefont {Stefan}\ \bibnamefont
  {Sandfeld}}, \bibinfo {author} {\bibfnamefont {Zoe}\ \bibnamefont
  {Budrikis}}, \bibinfo {author} {\bibfnamefont {Stefano}\ \bibnamefont
  {Zapperi}}, \ and\ \bibinfo {author} {\bibfnamefont {David~Fernandez}\
  \bibnamefont {Castellanos}},\ }\bibfield  {title} {\enquote {\bibinfo {title}
  {Avalanches, loading and finite size effects in 2d amorphous plasticity:
  results from a finite element model},}\ }\href@noop {} {\bibfield  {journal}
  {\bibinfo  {journal} {Journal of Statistical Mechanics: Theory and
  Experiment}\ }\textbf {\bibinfo {volume} {2015}},\ \bibinfo {pages} {P02011}
  (\bibinfo {year} {2015})}\BibitemShut {NoStop}%
\bibitem [{\citenamefont {Lin}\ and\ \citenamefont
  {Wyart}(2016)}]{lin2016mean}%
  \BibitemOpen
  \bibfield  {author} {\bibinfo {author} {\bibfnamefont {Jie}\ \bibnamefont
  {Lin}}\ and\ \bibinfo {author} {\bibfnamefont {Matthieu}\ \bibnamefont
  {Wyart}},\ }\bibfield  {title} {\enquote {\bibinfo {title} {Mean-field
  description of plastic flow in amorphous solids},}\ }\href@noop {} {\bibfield
   {journal} {\bibinfo  {journal} {Physical review X}\ }\textbf {\bibinfo
  {volume} {6}},\ \bibinfo {pages} {011005} (\bibinfo {year}
  {2016})}\BibitemShut {NoStop}%
\bibitem [{\citenamefont {Jagla}(2017)}]{jagla2017different}%
  \BibitemOpen
  \bibfield  {author} {\bibinfo {author} {\bibfnamefont {Eduardo~Alberto}\
  \bibnamefont {Jagla}},\ }\bibfield  {title} {\enquote {\bibinfo {title}
  {Different universality classes at the yielding transition of amorphous
  systems},}\ }\href@noop {} {\bibfield  {journal} {\bibinfo  {journal}
  {Physical Review E}\ }\textbf {\bibinfo {volume} {96}},\ \bibinfo {pages}
  {023006} (\bibinfo {year} {2017})}\BibitemShut {NoStop}%
\bibitem [{\citenamefont {Bi}\ \emph {et~al.}(2015)\citenamefont {Bi},
  \citenamefont {Lopez}, \citenamefont {Schwarz},\ and\ \citenamefont
  {Manning}}]{Bi2015}%
  \BibitemOpen
  \bibfield  {author} {\bibinfo {author} {\bibfnamefont {Dapeng}\ \bibnamefont
  {Bi}}, \bibinfo {author} {\bibfnamefont {JH}~\bibnamefont {Lopez}}, \bibinfo
  {author} {\bibfnamefont {Jennifer~M}\ \bibnamefont {Schwarz}}, \ and\
  \bibinfo {author} {\bibfnamefont {M~Lisa}\ \bibnamefont {Manning}},\
  }\bibfield  {title} {\enquote {\bibinfo {title} {A density-independent
  rigidity transition in biological tissues},}\ }\href@noop {} {\bibfield
  {journal} {\bibinfo  {journal} {Nature Physics}\ }\textbf {\bibinfo {volume}
  {11}},\ \bibinfo {pages} {1074--1079} (\bibinfo {year} {2015})}\BibitemShut
  {NoStop}%
\end{thebibliography}%

\end{document}


\title{Supplemental Material\\Random traction yielding transition in epithelial tissues}

\author{Aboutaleb Amiri}
\affiliation{Max Planck Institute for the Physics of Complex Systems, N\"othnitzer Str. 38,
01187 Dresden, Germany.}

\author{Charlie Duclut}
\affiliation{Max Planck Institute for the Physics of Complex Systems, N\"othnitzer Str. 38,
01187 Dresden, Germany.}
\affiliation{Universit\'e Paris Cit\'e, Laboratoire Mati\`ere et Syst\`emes Complexes (MSC), UMR 7057 CNRS, Paris,  France}
\affiliation{Laboratoire Physico-Chimie Curie, CNRS UMR 168, Institut Curie, Universit\'e PSL, Sorbonne Universit\'e, 75005, Paris, France}
\author{Frank J\"ulicher}\email{julicher@pks.mpg.de}
\affiliation{Max Planck Institute for the Physics of Complex Systems, N\"othnitzer Str. 38,
01187 Dresden, Germany.}
\affiliation{Cluster of Excellence Physics of Life, TU Dresden, Dresden, Germany}
\affiliation{Center for Systems Biology Dresden, Dresden, Germany}
\author{Marko Popovi\'c}\email{mpopovic@pks.mpg.de}
\affiliation{Max Planck Institute for the Physics of Complex Systems, N\"othnitzer Str. 38,
01187 Dresden, Germany.}
\affiliation{Center for Systems Biology Dresden, Dresden, Germany}

\date{\today}
\maketitle

\onecolumngrid

\section{Confinement of vertex model on spherical geometry}\label{App:confinement}
We consider a non-deforming spherical geometry by setting ${\bm u}_m\cdot \hat{n}_m=0$ in the force balance Eq.~(1) in the main text, where $\hat{n}_m$ is the normal vector to sphere at vertex $m$. This leads to the definition of the normal force ${\bm f}_m^n = f_m^n \hat{n}_m$ at vertex $m$ with the magnitude:
\begin{equation}
    f_m^n = \left[ \frac{\partial W}{\partial {\bm X}_m} - {\bm f}_m^a\right] \cdot \hat{n}_m \quad.
\end{equation}

\section{Initial Conditions}
We initialize the tissue configuration by Voronoi diagram construction of $N$ randomly distributed cell centers on a sphere of radius $R=(N A_0/4\pi)^{1/2}$, where $A_0$ is the cell preferred area. We initialize the cell polarity vectors ${\bm p}_\alpha$ in tangential plane of the spherical tissue, and with the random direction angle from a uniform distribution (see Fig.~1a in the main text).

\section{Triangulation of vertex model tissue}\label{App:VM}
\begin{figure}[b]
    \centering
    \includegraphics[width=0.5\linewidth]{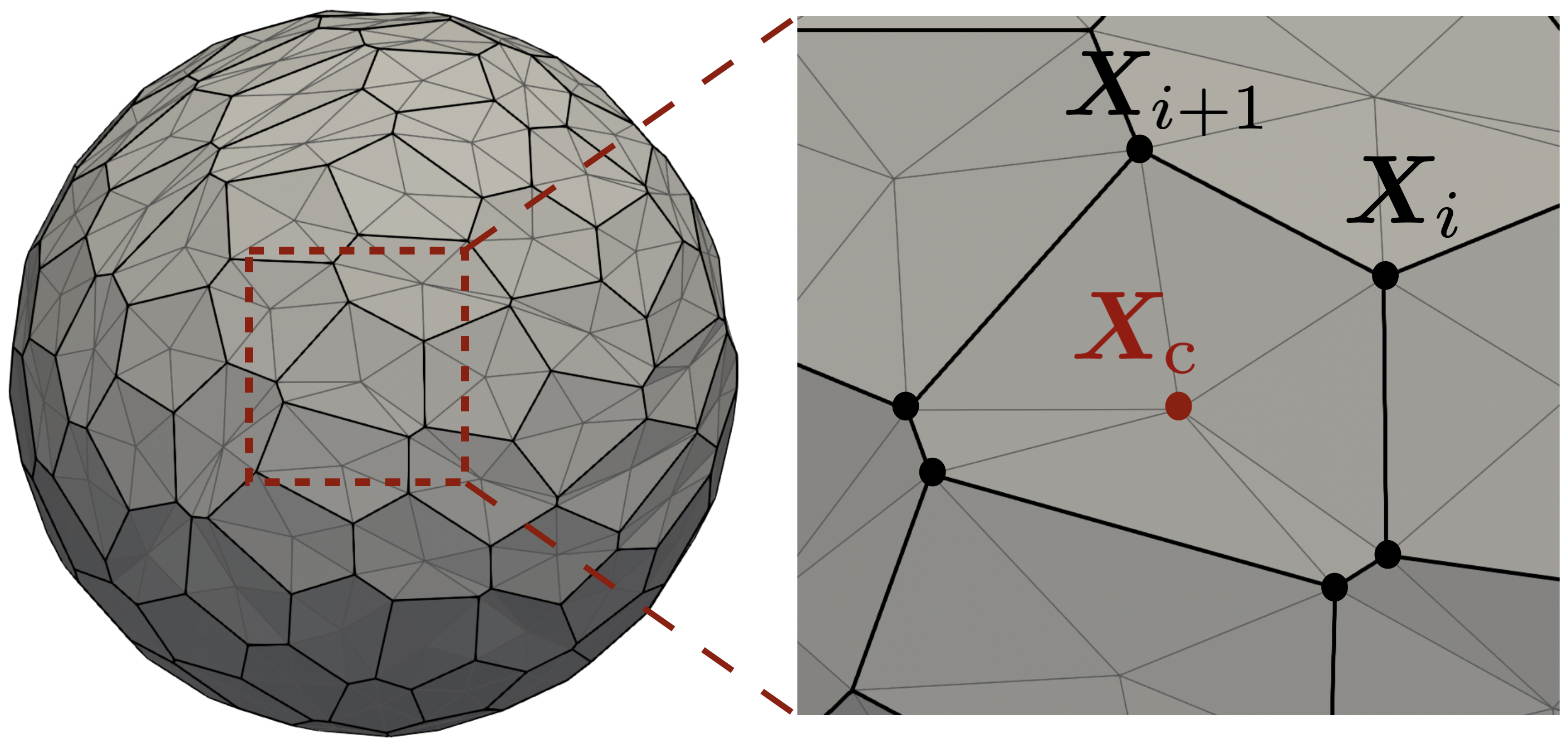}
    \caption{Vertex model triangulation by connecting consecutive cell vertices to its center.}
    \label{fig:App:VM}
    \end{figure}

In our vertex model on 2D sphere, cell vertices in general are not co-planar. Therefore, a unique definition of cell geometric quantities such as cell area requires a triangulation definition. Here, we construct the triangulation, as depicted in Fig.~\ref{fig:App:VM}, by connecting consecutive cell vertices ${\bm X_i}$ and ${\bm X_{i+1}}$, and the cell centroid given by
\begin{equation}
    {\bm X}_\text{c} = \frac{1}{M} \sum_{i=1}^{M} {\bm X}_i,
\end{equation}
where $M$ denotes number of cell vertices. Consequently, the cell area reads
\begin{equation}
    A = \frac{1}{2} \sum_{i=1}^M |({\bm X}_i - {\bm X}_\text{c})\times ({\bm X}_{i+1} - {\bm X}_\text{c})|,
\end{equation}
with the consideration ${\bm X_{M+1}}\equiv {\bm X_1}$.

\section{Implementation of convective co-rotational time derivative}
\label{DDt}
We initialize the cell polarities in tangential plane of the spherical tissue and in random directions  from a uniform distribution. They are transported by cells carrying them
through the convective co-rotational time derivative $D{\bm p}_\alpha/Dt$ in Eq.~(3) in the main text, that can be written in a discrete form
\begin{align}
    &{\bm p}_\alpha(t+\Delta t) = \underline{\underline{R}}({\bf \Omega}_\alpha \Delta t) {\bm p}_\alpha(t),\\
    &|{\bm p}_\alpha(t+\Delta t)|=1,
\end{align}
where the three dimensional rotation matrix $\underline{\underline{R}}({\bf \Omega}_\alpha \Delta t)$ is constructed by extracting the solid body angular velocity of each cell ${\bf \Omega}_\alpha$ based on the velocity of its vertices 
\begin{align}
    {\bm u}_{m,\alpha}(t) =& \frac{{\bm X}_{m,\alpha}(t+\Delta t)-{\bm X}_{m,\alpha}(t)}{\Delta t},\\
    {\bm u}_{m,\alpha}(t) =& \boldsymbol{\Omega}_\alpha(t)\times {\bm r}_{m,\alpha}(t) + {\bm u}_\alpha(t) + \delta {\bm u}_{m,\alpha}(t),
    \label{vel_decomposition}
\end{align}
where ${\bf X}_{m,\alpha}(t)$ is the position of each vertex $m$ that belongs to cell $\alpha$, and ${\bm r}_{m,\alpha}(t) = {\bm X}_{m,\alpha}(t)-{\bm X}_{\alpha, {\rm c}}$. Cell centroid and its translational velocity are defined by
\begin{align*}
    {\bm X}_{\alpha, {\rm c}}(t) =& \frac{1}{M_\alpha}\sum_{m\in\alpha} {\bm X}_{m,\alpha}(t),\\
    {\bm u}_\alpha(t) =& \frac{1}{M_\alpha}\sum_{m\in\alpha} {\bm u}_{m,\alpha}(t),
\end{align*}
and $M_\alpha$ is the number of vertices of cell $\alpha$. The last term on the right-hand side of Eq.~(\ref{vel_decomposition}), $\delta {\bm u}_{m,\alpha}(t)$, is the residual velocity of each cell vertex after subtraction of the cell solid body rotation.

In order to compute $\boldsymbol{\Omega}_\alpha(t)$, we define the angular momentum $\boldsymbol{\Gamma}_{m,\alpha}(t)$ of vertex $m$ as:
\begin{align}
\label{eq_momentum1}
    \boldsymbol{\Gamma}_{m,\alpha}(t) = {\bm r}_{m,\alpha}(t)\times {\bm u}_{m,\alpha}(t) - {\bm r}_{m,\alpha}(t) \times {\bm u}_\alpha(t) \, .
\end{align}
Inserting this definition into Eq.~\eqref{vel_decomposition} yields:
\begin{align}
\label{eq_angMomentum1}
    \boldsymbol{\Gamma}_{m,\alpha} = \boldsymbol{\Omega}_\alpha {\bm r}_{m,\alpha}^2 - {\bm r}_{m,\alpha}\left[ \boldsymbol{\Omega}_\alpha \cdot {\bm r}_{m,\alpha} \right] + {\bm r}_{m,\alpha} \times \delta {\bm u}_{m,\alpha} \, ,
\end{align}
where the time dependence has been omitted for simplicity. Equation~\eqref{eq_angMomentum1} can then be rewritten in matrix form as:

\begin{align}
    \boldsymbol{\Gamma}_{m,\alpha} = \boldsymbol{M}_{m,\alpha} \cdot \boldsymbol{\Omega}_\alpha + {\bm r}_{m,\alpha}\times \delta {\bm u}_{m,\alpha} \, ,
\end{align}
where we have introduced the moment of inertia tensor $\boldsymbol{M}_{m,\alpha}$ of vertex $m$:
\begin{align}
    \boldsymbol{M}_{m,\alpha} = {\bm r}_{m,\alpha}^2 \mathbbm{1} - {\bm r}_{m,\alpha}^T {\bm r}_{m,\alpha}\, .
\end{align}
The average angular momentum of cell $\alpha$ is obtained as:
\begin{align}
\label{eq_AvgGamma1}
    \boldsymbol{\Gamma}_\alpha = \frac{1}{N}\sum_{m\in\alpha} \boldsymbol{\Gamma}_{m,\alpha} = \boldsymbol{M}_\alpha \cdot \boldsymbol{\Omega}_\alpha + \frac{1}{M_\alpha} \sum_{m\in\alpha} {\bm r}_{m,\alpha}\times \delta {\bm u}_{m,\alpha} \, ,
\end{align}
where $\boldsymbol{M}_\alpha=\sum_{m\in\alpha} \boldsymbol{M}_{m,\alpha}/M_\alpha$ is the average moment of inertia tensor. In the case of a solid-body rotation, $\delta {\bm u}_{m,\alpha}(t)=\bm 0$, such that the angular velocity is simply obtained from Eq.~\eqref{eq_AvgGamma1} as:
\begin{align}
    &\boldsymbol{\Omega}_\alpha = \boldsymbol{M}_\alpha^{-1} \cdot \boldsymbol{\Gamma}_0^\alpha \, ,
\end{align}

with $\boldsymbol{\Gamma}_0^\alpha = \sum_{m\in\alpha} ({\bm X}_{m,\alpha} - {\bm X}_{\alpha, {\rm c}}) \times ({\bm u}_{m,\alpha} - {\bm u}_\alpha)/M_\alpha$.\newline

\section{Uncertainties in detecting avalanches}\label{App:Skip_Interval}

Duration of an avalanche can be uniquely defined in the quasistatic limit $v \to 0$ as the time period during which $F$ is decreasing due to successive cell rearrangements. After the avalanche no cell rearrangements occur until $F$ in the system increases sufficiently by spring displacement to trigger the next cell rearrangements. Duration of this loading period diverges in the limit $v \to 0$ and avalanches can be precisely identified.

At a finite $v$, the observed time intervals during which $F$ is decreasing depend on the time resolution $\delta t$ at which the data is recorded. Each spring moves a distance $\delta s = v \delta t$ in a time interval $\delta t$ (see  Eq.~5 in the main text) during which $F$ increases by $k \delta s$ due to spring movement. Occasionally during an avalanche the $F$ decrease rate can fall below $k v$ and the $F$ slightly increases although the avalanche is still ongoing. If the time resolution $\delta t$ is very small these slight increases will often be recorded effectively splitting original avalanches into smaller ones. On the other hand choosing too large $\delta t$ leads to merging of $F$ decrease intervals belonging to different avalanches. 

\begin{figure}
    \centering
    \includegraphics[width=0.95\linewidth]{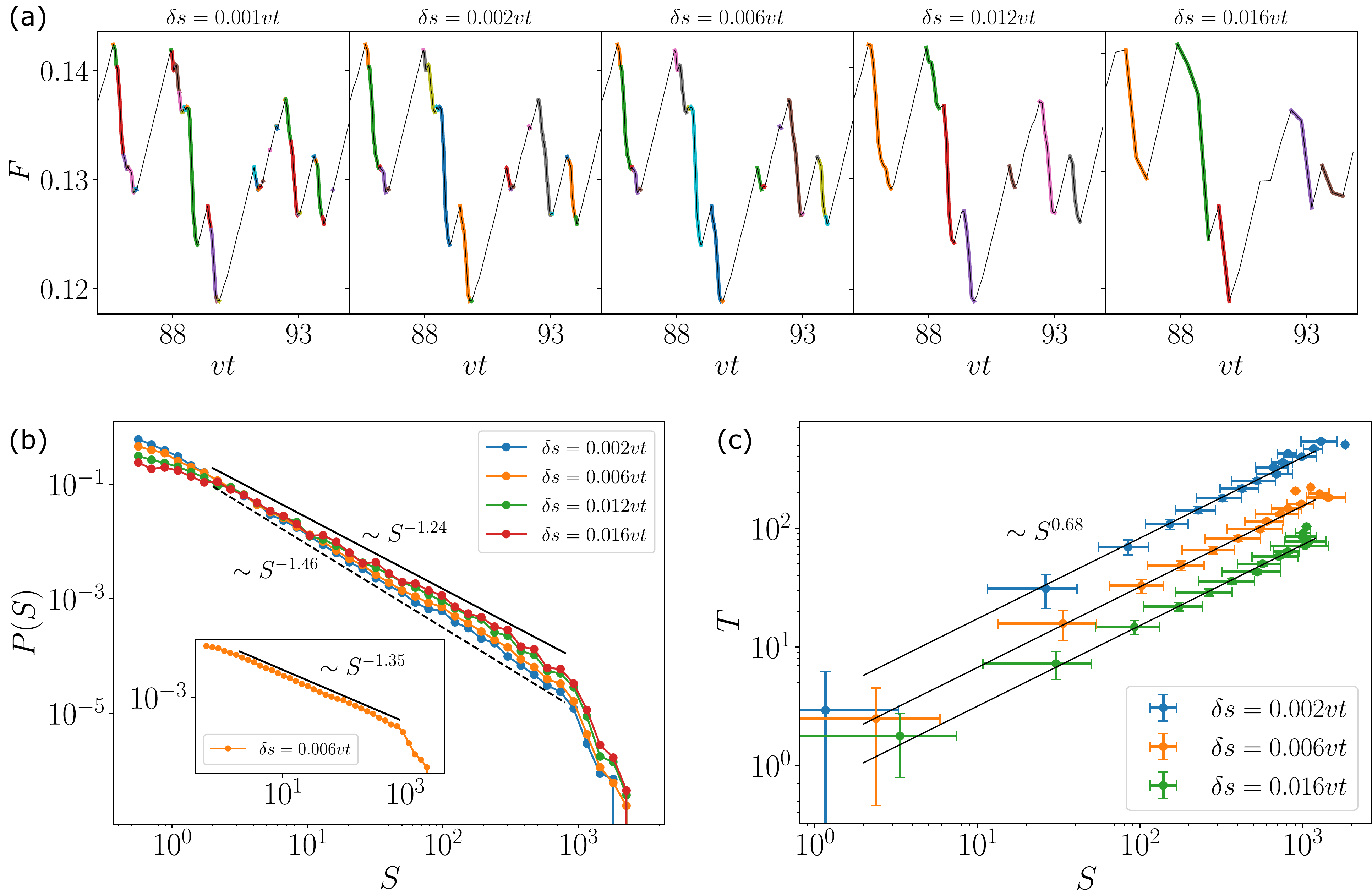}
    \caption{Avalanche statistics. \textbf{(a)}~Uncertainty in detecting drops in the average traction force magnitude due to the finite time resolution. \textbf{(b)}~Measured avalanche size distribution as a function of time resolution $\delta t$ allows us to estimate the uncertainty of the avalanche distribution exponent $\tau=1.35\pm 0.11$. \textbf{(c)}~Varying the time resolution does not change  avalanche duration scaling with its size. Blacl lines indicate $T\sim S^{0.68}$.}
    \label{fig:SI:T_S}
\end{figure}

We first determine the extreme limits of low and high $\delta t$ for which described artifacts leading to splitting and merging of $F$ decrease intervals are clearly visible (see Fig.~\ref{fig:SI:T_S}a). We then estimate the uncertainty in our measurements of avalanche size distribution by varying the time resolution $\delta t$ between these extreme limits and identify the intermediate value of $\delta t$ for which we test the robustness of the results upon further decreasing $v$ (see Fig.~\ref{fig:SI:T_S}b). 

Note that varying the time resolution $\delta t$ does not change avalanche duration scaling with its size (see Fig.~\ref{fig:SI:T_S}c).

\section{Fractal dimension of avalanches}
\label{App:Fractal_Dim}
\begin{figure}
    \centering
    \includegraphics[width=0.9\linewidth]{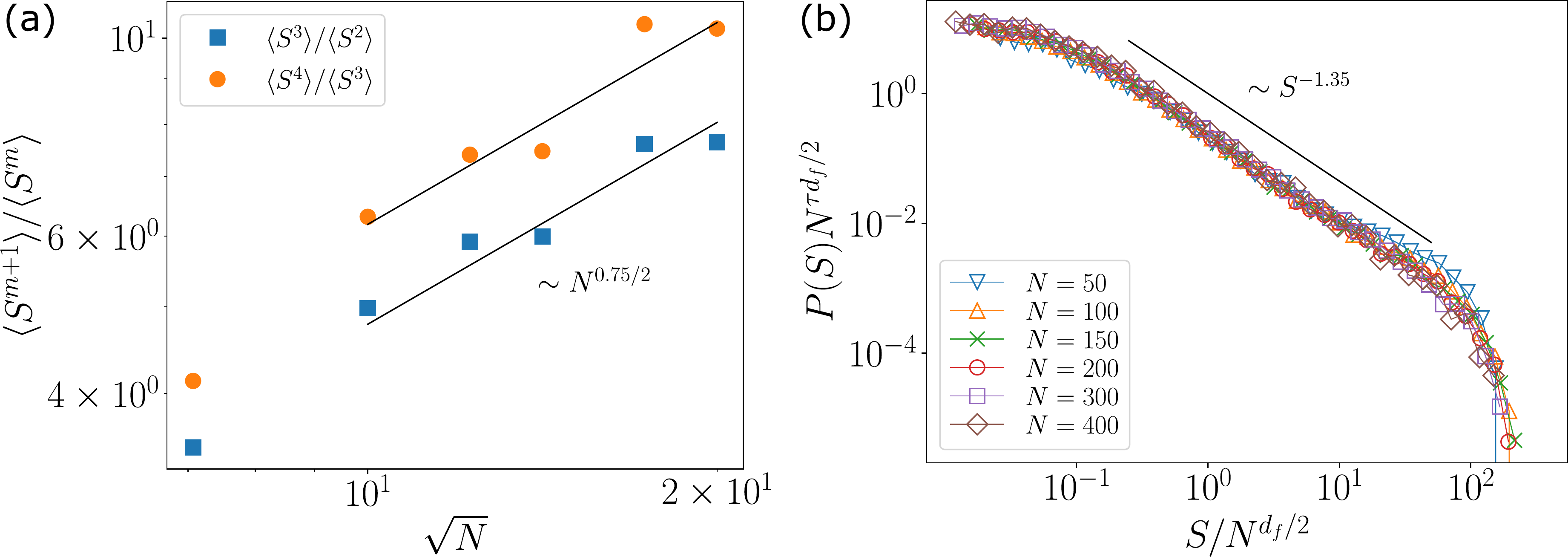}
    \caption{Avalanche statistics. \textbf{(a)}~Ratio of consecutive moments of avalanche size distribution scales with linear system size $\sqrt{N}$ as $\langle S^{m+1}\rangle/\langle S^{m}\rangle \sim N^{d_f/2}$, with $d_f=0.75\pm0.15$. \textbf{(b)}~Collapse of avalanche size distribution for different system sizes with the exponents $\tau=1.35$, and $d_f=0.75$.}
    \label{fig:SI:df}
\end{figure}

The $m$-th moment of avalanche size distribution $P(S)$ reads
\begin{equation}
    \langle S^{m} \rangle = \int_1^{S_c} S^m P(S) {\rm d}S,
\end{equation}
where $S_c$ is the cut-off size of the avalanches. Considering a power-law normalized avalanche size distribution $P(S)\sim S^{-\tau}$ (Fig.~2a in the main text), it follows
\begin{equation}
    \frac{\langle S^{m+1}\rangle}{\langle S^m\rangle} \sim S_c.
\end{equation}
The system size dependent cut-off value in the normalized avalanche size distribution (Fig.~2a in the main text) scales with linear system size as $S_c\sim N^{d_f/d}$, where $d_f$ is the fractal dimension of avalanches, and $d=2$ is  the system dimension. The fit (shown by black lines in Fig.~\ref{fig:SI:df}a) yields $d_f=0.75\pm 0.15$.

Taking the estimation of $d_f\approx0.75$ and $\tau\approx1.35$ (see main text) leads to a collapse of the tail in the avalanche size distribution for various system sizes as is shown in Fig.~\ref{fig:SI:df}b.

\section{Density of plastic excitations}
\begin{figure}
    \centering
    \includegraphics[width=0.8\linewidth]{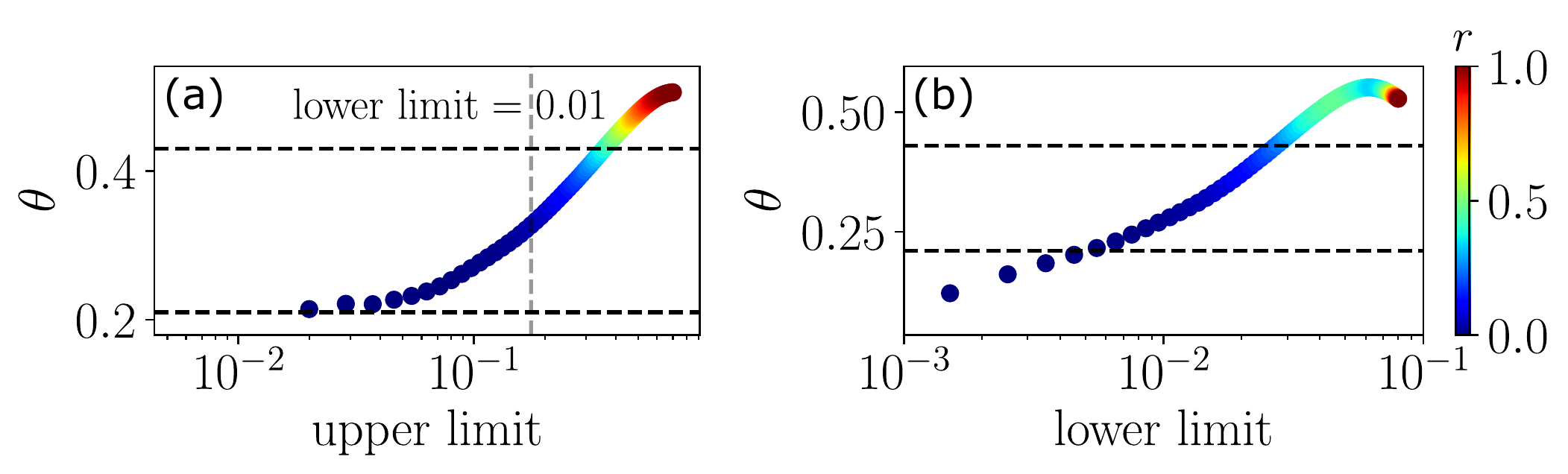}
    \caption{The fit of bond length distribution to power-law $P(\ell)\sim \ell^\theta$, by considering various values of the lower and upper limit of bond lengths for the fit. The horizontal dashes lines show the range of $\theta$ value predicted by the scaling relation (Eq.~6 in the main text). \textbf{(a)} Shows the measured values of $\theta$ for different values of upper limit while fixing the lower limit at 0.01. The vertical line shows the upper limit value that is shown in Fig.~3b of the main text. \textbf{(b)} Shows how the measurement of $\theta$ changes as we shift the fitting window by varying the lower limit and doing the fit for one decade. The color shows $r=(1/K)\sum_{j=0}^K (P(\ell_j)-\ell_j^{\theta})^2$, the average of squared residuals of the least squares fit, where sum is over $K$ bonds with lengths in the fitting interval.}
    \label{fig:tta_grid}
\end{figure}

In our vertex model, we determine the exponent $\theta$ by fitting cumulative bond length distribution (see Fig.~3b in the main text) for a wide range, and compare the measured values with the range predicted by the scaling relation (main text Eq.~6).
Further analysis (Fig.~\ref{fig:tta_grid}) shows that varying lower and upper limits of the fitting range leads to measurements of $\theta$ that vary in a range $[0.21, 0.52]$. We find that both increasing the upper limits while fixing the lower limit (Fig.~\ref{fig:tta_grid}a) and shifting up the one decade-long fitting interval (Fig.~\ref{fig:tta_grid}b) lead to an increase in our measurement of $\theta$. To test the quality of each measurement, we quantify the average of squared residuals of the fit $r=(1/K)\sum_{j=0}^K (P(\ell_j)-\ell_j^{\theta})^2$, where sum is over $K$ bonds with lengths in the fitting interval. We find that the confidence of the fit reduces as the measured values goes above the predicted range, marked by dashed lines in Fig.~\ref{fig:tta_grid}.

\section{Critical traction force magnitude}\label{App:F_c}

Figure~4 in the main text suggests that the magnitude of critical traction force $F_c(N)$ is system-size dependent and is expected to be of the form $F_c(N) - F_c \sim  N^{-1/(d \nu)}$, since a system smaller than a correlation length $\xi \sim (F - F_c)^{-\nu}$ cannot be distinguished from the system at $F_c$. 
Fitting the Herschel-Bulkley law to numerical simulation data (Fig.~4 in the main text) reveals $F_c(100)\approx0.128$ and $F_c(200)\approx0.119$. The decrease of $F_c(N)$ is qualitatively consistent with our expectation, however, an investigation of $F_c$ in a broader range of system sizes would be required to test the scaling prediction and the value of correlation length exponent $\nu= 0.8 \pm 0.096$ obtained from the scaling relation $\nu= 1/(d - d_f)$. 

\section{Random Yielding transition in flat bi-periodic vertex model}\label{App:flat_biperiodic}
\begin{figure}[htb]
	\centering
    \includegraphics[width=0.8\linewidth]{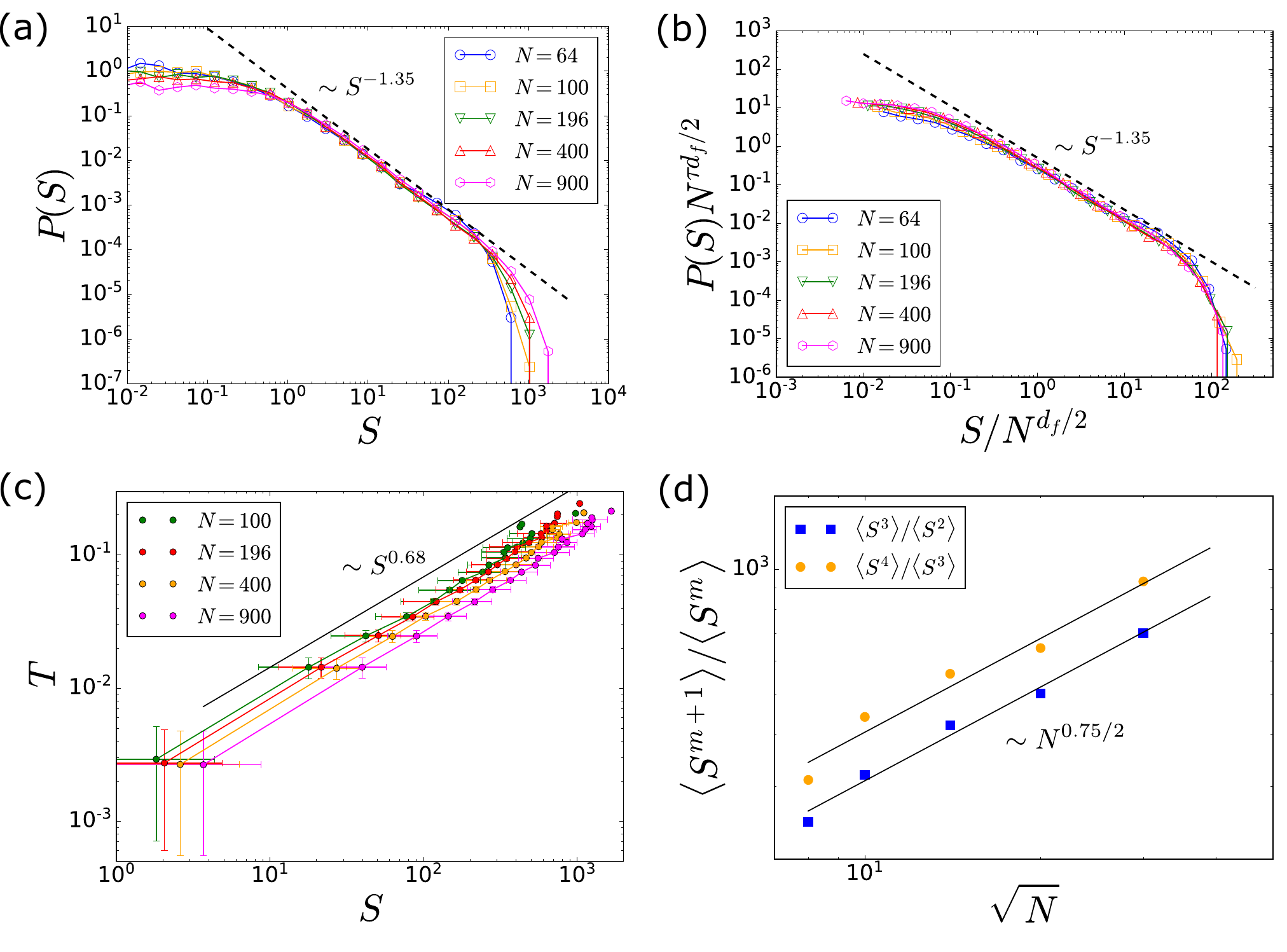}
	\caption{
    Avalanche statistics in flat bi-periodic vertex model tissues. (a) avalanche size distribution $P(S)$ is a power law with exponent $\tau\approx1.35$ indicated by the dashed line. (b) Collapse of $P(S)$ with $d_f=0.75$ for a wide range of tissue sizes. (c) Avalanche duration \emph{vs} avalanche size. The solid black line shows $T\sim S^{z/d_f}$, with $z/d_f\approx0.68$. (d) Ratio of consecutive moments of avalanche size distribution scales with linear system size $\sqrt{N}$ as $\langle S^{m+1}\rangle/\langle S^{m}\rangle \sim N^{d_f/2}$, with $d_f=0.75$ marked by the solid black lines.}
	\label{fig:flat_biperiodic}
\end{figure}

In order to test whether the spherical geometry affects the values of the critical exponents, we have measured the exponents $\tau$, $z$ and $d_f$ in a flat two-dimensional vertex model with bi-periodic boundary conditions. We find a power-law avalanche size distribution consistent with the exponent $\tau\approx1.35$ we reported for the spherical geometry (Fig.~\ref{fig:flat_biperiodic}a). Moreover, the  fractal dimension $d_f\approx0.75$ measured in the spherical geometry leads to the collapse of the avalanche size distribution (see Fig.~\ref{fig:flat_biperiodic}b) and the same value is consistent with the finite-size scaling of the consecutive moments of avalanche size statistics (Fig.~\ref{fig:flat_biperiodic}d). Finally, we find that avalanche duration scales with avalanche size with an exponent $z/d_f\approx0.68$, consistent with the value of exponent $z$ reported in the spherical geometry (Fig.~\ref{fig:flat_biperiodic}c). 

